\documentclass[pdflatex,sn-nature]{sn-jnl}
\usepackage{graphicx}
\usepackage{multirow}
\usepackage{amsmath,amssymb,amsfonts}
\usepackage{amsthm}
\usepackage{mathrsfs}
\usepackage[title]{appendix}
\usepackage{xcolor}
\usepackage{textcomp}
\usepackage{manyfoot}
\usepackage{booktabs}
\usepackage{algorithm}
\usepackage{algorithmicx}
\usepackage{algpseudocode}
\usepackage{listings}
\usepackage{soul}
\usepackage{xr}
\usepackage{geometry}
\usepackage{placeins}
\theoremstyle{thmstyleone}

\theoremstyle{thmstyletwo}

\theoremstyle{thmstylethree}

\begin{document}

\title[Article Title]{Generative deep learning improves reconstruction of global historical climate records}

\author[1,2]{\fnm{Zhen} \sur{Qian}}
\author*[1,2,3]{\fnm{Teng} \sur{Liu}}\email{teng.liu@tum.de}
\author[4]{\fnm{Duo} \sur{Chan}}
\author[1,2]{\fnm{Sebastian} \sur{Bathiany}}
\author[1,2]{\fnm{Shangshang} \sur{Yang}}
\author[1,2]{\fnm{Philipp} \sur{Hess}}
\author[2,5]{\fnm{Nils} \sur{Bochow}}
\author[1]{\fnm{Christian} \sur{Burmester}}
\author[1,2]{\fnm{Maximilian} \sur{Gelbrecht}}
\author[2]{\fnm{Brian} \sur{Groenke}}
\author*[1,2,6]{\fnm{Niklas} \sur{Boers}}\email{n.boers@tum.de}

\affil[1]{\orgdiv{Munich Climate Center and Earth System Modelling Group}, \orgname{Department of Aerospace and Geodesy, TUM School of Engineering and Design, Technical University of Munich}, \city{Munich}, \country{Germany}}
\affil[2]{\orgname{Potsdam Institute for Climate Impact Research}, \city{Potsdam}, \country{Germany}}
\affil[3]{\orgdiv{School of Systems Science and Institute of Nonequilibrium Systems}, \orgname{Beijing Normal University}, \city{Beijing}, \country{China}}
\affil[4]{\orgdiv{School of Ocean and Earth Science}, \orgname{University of Southampton}, \city{Southampton}, \country{UK}}
\affil[5]{\orgdiv{Alfred Wegener Institute}, \orgname{Helmholtz Centre for Polar and Marine Research}, \city{Potsdam}, \country{Germany}}
\affil[6]{\orgdiv{Department of Mathematics and Global Systems Institute}, \orgname{University of Exeter}, \city{Exeter}, \country{UK}}

\abstract{
Instrumental observations from the early twentieth century are sparse and uncertain, limiting estimates of historical climate variability and change~\cite{morice2021updated,harris2020version}. 
Existing reconstructions rely largely on statistical interpolation~\cite{yamamoto2005correcting,rohde2013new}, smoothing spatiotemporal variability and leading to underestimation of local extremes.
Recent deterministic machine-learning approaches focus primarily on spatial inpainting of coarse-resolution temperature fields but do not explicitly quantify uncertainty or enforce spatiotemporal consistency~\cite{kadow2020artificial,bochow2025reconstructing}.
Here, we present GenCRest, a generative deep-learning model that reconstructs globally complete temperature and precipitation fields from sparse, uncertain observations using a generative prior learned from Earth-system models and reanalyses.
GenCRest preserves accuracy, temporal persistence, and higher-order variability of climate dynamics, consistently outperforming statistical and machine-learning benchmarks.
Applying GenCRest to observational records that underpin climate assessments, we find stronger early-twentieth-century El Ni\~no--Southern Oscillation (ENSO) variability and identify approximately 50\% more El Ni\~no events documented in the literature, which are not represented in established datasets.
Our reconstructions reveal a more spatially heterogeneous history of temperature change during 1900--1980, with pronounced regional warming and cooling contrasts, particularly in the Arctic.
For 1981--2024, our results suggest that recent assessments overestimate Arctic-mean warming due to interpolation smoothing, while underestimating warming in the Greenland, Barents, and Kara Seas.
Furthermore, based on our seamless global reconstruction that infers precipitation variability across the oceans and under-monitored regions, we identify more widespread wet extremes over land in the early twentieth century and intensified regional wetting and drying across land and ocean in recent decades.
Our results provide a globally complete, uncertainty-aware, spatiotemporally consistent temperature and precipitation reconstruction back to 1850 and 1901, respectively, enabling more precise assessments of historical anthropogenic climate change, especially with regard to historical changes in extreme weather events, even in the polar regions where observations are sparse.
}

\maketitle

\section{Introduction}

Understanding historical climate change is essential for constraining the present state of the Earth system and narrowing uncertainty in future projections~\cite{tokarska2020past,gulev2021changing}.
However, the historical record of climate change remains incomplete.
Although some individual records extend further back, observations suitable for global analyses of near-surface temperature and precipitation become broadly available only from the nineteenth century on and remain particularly sparse over the oceans, polar regions, and many remote land areas~\cite{schneider2014gpcc,kennedy2019ensemble,morice2021updated,rohde2013new,harris2020version}.
Satellite observations have substantially expanded spatial coverage since the 1970s, but their record is comparatively short.
Many variables and regions continue to depend on heterogeneous and incomplete in situ observing networks~\cite{adler2003version,kent2021historical}.
Accurate knowledge of the evolution of temperature and precipitation is fundamental to characterizing Earth’s energy and water cycles~\cite{trenberth2007estimates,trenberth2009earth,stephens2012update}, long-term temperature trends, as well as past changes of climate variability and extreme events.
Observational gaps nevertheless obscure the pace and spatial structure of long-term warming, the dynamics of climate variability, and the occurrence of past extremes~\cite{cowtan2014coverage}.
Climate field reconstruction therefore has the potential to transform sparse, irregular observations into spatially and temporally coherent records, thereby supporting assessments of climate variability and change and their link to energy and water-cycle processes.

A range of methods has been developed for climate-field reconstruction, including distance-weighted interpolation, spatial and spatiotemporal Kriging, Gaussian process regression, optimal interpolation, and reduced-rank covariance methods~\cite{jones1997estimating,kaplan1998analyses,rohde2013new,morice2021updated,calvert2024improving,ishii2025historical,chan2026dcent}.
Methods from these broad categories have been used to construct widely established gridded observational datasets that support assessments of historical climate change, including those synthesized in the Intergovernmental Panel on Climate Change (IPCC)  Assessment Reports~\cite{gulev2021changing}.
These methods generally infer unobserved values from weighted statistical combinations of available observations based on spatial distance or covariance.
This can suppress spatial heterogeneity through smoothing~\cite{kaplan1998analyses,hofstra2009testing,risser2019probabilistic}, dampening local extremes and blurring distinct climatic gradients~\cite{hofstra2009testing,risser2019probabilistic,cowtan2014coverage}.
Consequently, while robust for diagnosing large-scale mean states, conventional interpolation can constrain the dynamic range of reconstructed climate fields and provide conservative estimates of variability in poorly observed regions (e.g., oceans, polar regions, and many remote land areas).

More recently, deterministic machine-learning approaches have formulated climate reconstruction primarily as a spatial inpainting problem.
Artificial neural networks have been used to infer missing temperature fields from the values available in data-supported regions~\cite{yin2024noaaglobaltemp}.
In particular, the partial-convolution approach~\cite{kadow2020artificial} and the Fourier-domain approach~\cite{bochow2025reconstructing} can recover sharper and more spatially coherent structures than conventional interpolation.
These inpainting models treat retained observations as fixed conditions, without an explicit mechanism to account for their uncertainty or correct observational errors.
Observational noise can therefore be retained in the completed fields, distorting estimates of temporal persistence and variability.
Their deterministic mappings also yield a single fixed reconstruction for each input, without representing the range of plausible climate fields given coarse observations.
Moreover, existing applications have also focused mainly on relatively coarse-resolution temperature fields and generally reconstruct individual spatial snapshots without explicitly constraining the temporal coherence of the resulting climate record~\cite{kadow2020artificial,yin2024noaaglobaltemp,bochow2025reconstructing}.
Temporal consistency is essential for accurately representing statistical properties relevant to climate dynamics, including autocorrelation, variance, and the persistence of extremes, which form the basis for assessing climate change impacts and changes in Earth system resilience~\cite{scheffer2009early,boers2022theoretical,boers2025destabilization,brovkin2021past,rietkerk2025ambiguity,liu2026SA}.

Generative deep learning constitutes a fundamental departure in which reconstruction is treated as probabilistic inference over complete spatiotemporal trajectories conditioned on incomplete observations.
Generative approaches are increasingly used in weather and climate modeling to represent forecast uncertainty, emulate complex Earth-system variability, and generate spatially and temporally coherent fields~\cite{price2025probabilistic,hess2025fast,brenowitz2025climate}.
Diffusion models are particularly suited to this setting because they can be trained unconditionally to learn a distribution of spatiotemporally consistent climate fields, which serves as a Bayesian prior~\cite{song2020score,song2021solving,kawar2022denoising,chung2022diffusion,yang2025generative,aich2026wind}.
Conditioning on sparse, irregular observations while accounting for their uncertainty produces a posterior distribution that can be sampled using likelihood guidance~\cite{rozet2023score,li2024learning,manshausen2025generative}.
Repeated sampling from this posterior yields an ensemble of plausible climate histories, representing reconstruction uncertainty in sparsely observed regions.

Here, we present GenCRest, a generative diffusion-based model that reconstructs ensembles of globally complete, spatiotemporally consistent historical temperature and precipitation fields from sparse, uncertain instrumental observations.
It does so by conditioning on these observations using a likelihood that incorporates the observational-error covariance matrix, rather than filling gaps through statistical interpolation.
In a sense, GenCRest is akin to data assimilation: it infers the most likely physical system state given observations, but uses a generative diffusion model trained on diverse model and reanalysis datasets rather than a dynamical numerical model to represent the system state.
We demonstrate that GenCRest consistently improves temporal consistency over advanced interpolation and machine-learning benchmarks, while better preserving spatial structure and the higher-order statistics of non-Gaussian variability.
This fidelity, coupled with ensemble-based estimates of reconstruction uncertainty, provides a stronger basis for characterizing variability and extremes in sparsely observed regions, including the polar regions.
Applying GenCRest to records that underpin climate assessments, we find stronger early-twentieth-century El Ni\~no--Southern Oscillation (ENSO) variability along with more widespread wet extremes over land.
At longer timescales, our reconstructions reveal a more spatially heterogeneous history of Arctic warming and intensified regional wetting and drying across land and ocean in recent decades.

\section{GenCRest: Generative climate reconstruction}\label{sec2}

GenCRest combines a learned spatiotemporal climate prior with observation-conditioned sampling to infer complete historical temperature and precipitation fields from sparse, uncertain observations (Fig.~\ref{fig:flowchart}; see Methods and Supplementary Methods for details).
The prior is learned by training a diffusion model unconditionally on complete climate fields, beginning with pre-training on historical simulations from the Coupled Model Intercomparison Project Phase 6 (CMIP6)~\cite{eyring2016overview}.
Fine-tuning on the fifth-generation ECMWF reanalysis (ERA5)~\cite{hersbach2020era5} then refines this prior using an observation-constrained representation of the climate system (Fig.~\ref{fig:flowchart}a).
At inference, we condition the learned prior on the available observations through a likelihood that guides reverse-diffusion denoising.
This likelihood incorporates reported uncertainty estimates through an error covariance matrix, including correlations between observational errors where available.
Temporal guidance promotes agreement between overlapping segments, reducing discontinuities when they are combined into a long climate record (Fig.~\ref{fig:flowchart}b; see Methods and Supplementary Methods for details).

Conditional sampling can be tailored to specific scientific objectives and available computational resources by balancing the refinement of individual realizations against ensemble size.
We illustrate this flexibility with two sampling configurations of GenCRest: GenCRest-HF and GenCRest-LE.
Within each variable-resolution setting, they share the same trained diffusion model and observations but use different denoising-step counts, guidance weights, and ensemble sizes under a comparable total computational budget (Fig.~\ref{fig:flowchart}c).
GenCRest-HF devotes more denoising steps to fewer realizations, allowing more extensive refinement through likelihood-guided posterior sampling~\cite{li2024learning} and more faithful representation of regional spatial structure, temporal evolution, and higher-order variability.
GenCRest-LE uses shorter sampling trajectories to generate a larger ensemble, yielding a more stable ensemble mean and a larger sample for characterizing reconstruction uncertainty.
In the applications presented here, we use GenCRest-HF primarily for regional climate processes, variability, and extremes, and GenCRest-LE primarily for mean climate states and their uncertainty.

\begin{figure}[!htbp]
\centering
\includegraphics[width=\linewidth]{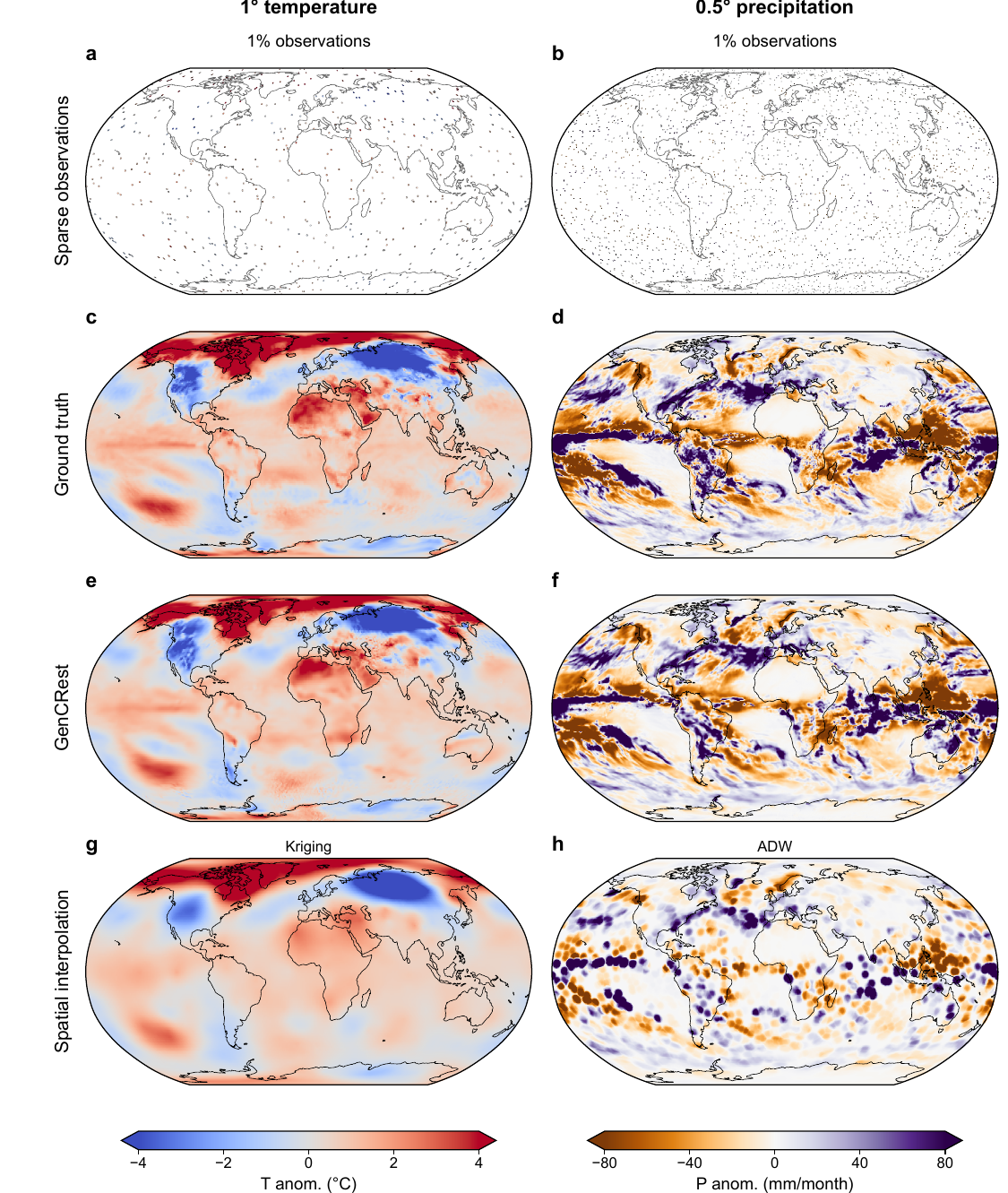}
\caption{
\textbf{Idealized, noise-free gap filling of temperature and precipitation anomalies.}
Held-out ERA5 fields for December 2009 are randomly subsampled at 1\% coverage, with retained values treated as exact observations.
Columns show $1^\circ$ temperature (\textbf{a}, \textbf{c}, \textbf{e}, \textbf{g}) and $0.5^\circ$ precipitation (\textbf{b}, \textbf{d}, \textbf{f}, \textbf{h}).
\textbf{a}, \textbf{b}, Sparse observations, with unobserved grid cells left blank.
\textbf{c}, \textbf{d}, Complete ERA5 reference fields.
\textbf{e}, \textbf{f}, GenCRest-HF reconstructions, each shown as a single realization.
\textbf{g}, \textbf{h}, Statistical interpolation using Kriging for temperature and angular-distance weighting (ADW) for precipitation.
Each column shares a common color scale across all four rows.
Further evaluations are provided in Extended Data Figs.~\ref{fig:temp5_random} and~\ref{fig:precip05_random} and Supplementary Sections~B.1 and~B.2.
}
\label{fig:random_demo}
\end{figure}

We examine GenCRest in an idealized setting by sparsifying ERA5 test fields to obtain uniformly random coverage over space and time, with the withheld values serving as evaluation references.
Reconstructions are performed at the resolutions of established gridded products, $5^\circ$ and $1^\circ$ for temperature and $0.5^\circ$ for precipitation, across a range of coverages (Fig.~\ref{fig:random_demo}; Extended Data Figs.~\ref{fig:temp5_random} and~\ref{fig:precip05_random}; Supplementary Sections~B.1 and~B.2).
As primary benchmarks, we use Kriging for temperature, which underlies the Berkeley Earth analysis~\cite{rohde2013new}, and angular-distance weighting (ADW) for precipitation, which is used in CRU TS gridding~\cite{harris2020version}.
We also compare with LaMa, a deep-learning inpainting approach recently adapted for historical temperature reconstruction~\cite{bochow2025reconstructing}, and ellipse Kriging, an advanced, spatially adaptive Kriging method used in the recent DCENT-I dataset~\cite{chan2026dcent}.

Our generative reconstructions preserve coherent regional variability and spatial detail (Fig.~\ref{fig:random_demo}; Extended Data Fig.~\ref{fig:precip05_random}).
In the $5^\circ$ temperature experiments, both GenCRest-HF and GenCRest-LE achieve higher temporal correlations and lower spatial errors across the tested observational coverages than the benchmarks, including additional machine-learning methods and a state-of-the-art Kriging variant (ellipse Kriging)~\cite{chan2026dcent} (Extended Data Fig.~\ref{fig:temp5_random}c; Supplementary Fig.~B4).
Kriging smooths finer-scale temperature features, whereas ADW produces radial patterns around isolated observations and attenuates the amplitude and small-scale variability of precipitation (Fig.~\ref{fig:random_demo}; Extended Data Fig.~\ref{fig:precip05_random}).
LaMa captures large-scale temperature patterns at $5^\circ$ but degrades markedly at higher resolution and for non-Gaussian precipitation, developing texture artifacts at $1^\circ$ for temperature and striping artifacts at $0.5^\circ$ for precipitation (Extended Data Fig.~\ref{fig:temp5_random}a; Supplementary Figs.~B5a and~B6).
These idealized experiments use noise-free observations with random uniform coverage over space and time, whereas historical records are unevenly distributed, change over time, and contain errors.

\section{Evaluation under realistic observational coverage and uncertainty}\label{sec3}

\begin{figure}[!htbp]
\centering
\includegraphics[width=.95\linewidth]{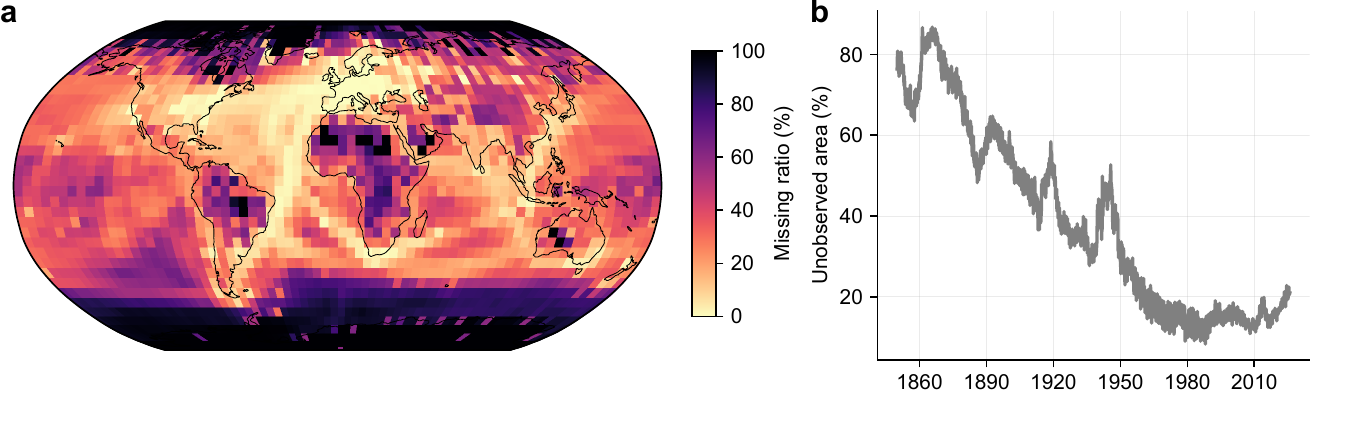}
\includegraphics[width=.95\linewidth]{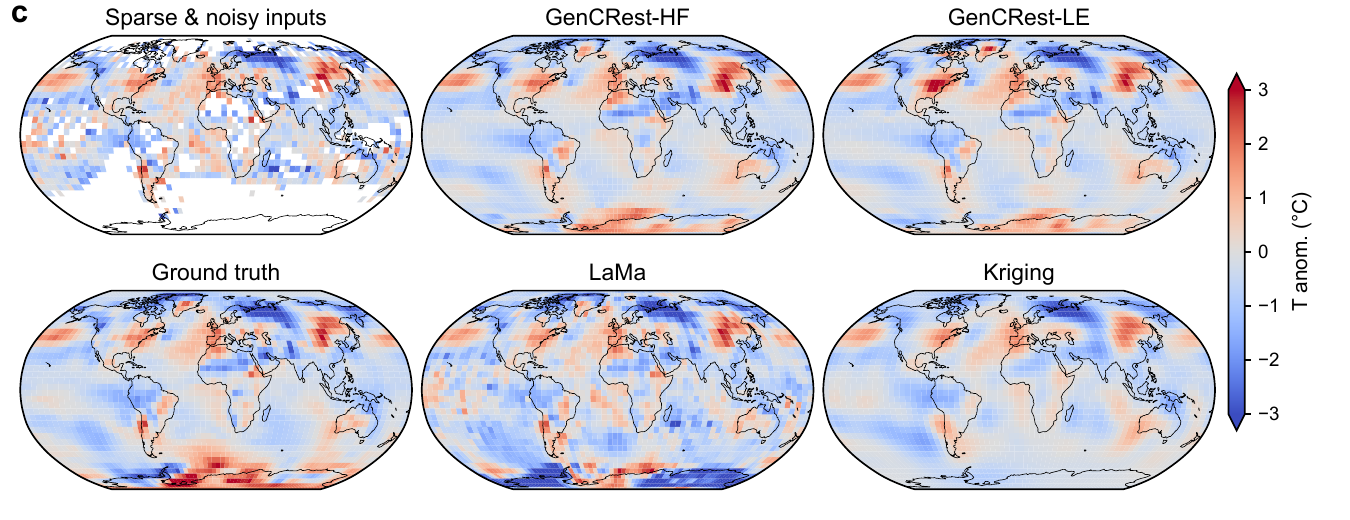}
\includegraphics[width=.95\linewidth]{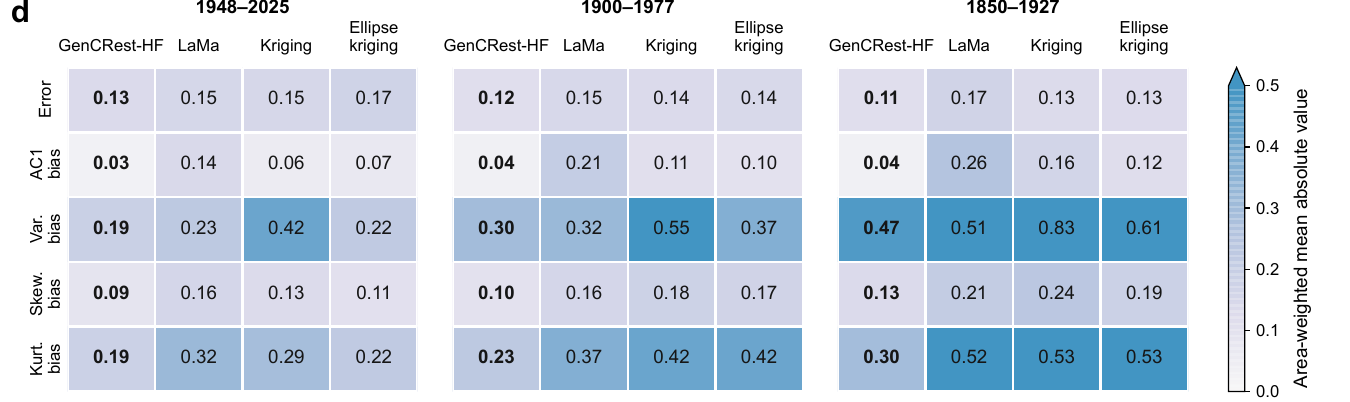}
\caption{
\textbf{Evaluation of temperature reconstruction under realistic observational coverage and uncertainty.}
\textbf{a}, Percentage of missing months in HadCRUT5 over 1850--2025.
\textbf{b}, Evolution of the unobserved area fraction.
For \textbf{c} and \textbf{d}, $5^\circ$ JRA-3Q fields over 1948--2025 are sampled using monthly HadCRUT5 masks and perturbed with errors drawn from the corresponding observational-error covariance matrices.
All methods use identical pseudo-observations, generated by applying monthly HadCRUT5 masks to JRA-3Q fields and perturbing the retained values according to the corresponding observational-error covariance matrices.
The complete, unperturbed JRA-3Q fields serve as the reference.
\textbf{c}, A randomly selected example for July 1949 under the temporally corresponding 1948--2025 masks, comparing the pseudo-observations, the reference field, and reconstructions from GenCRest-HF, GenCRest-LE, LaMa, and Kriging.
GenCRest-HF and GenCRest-LE maps show ensemble means.
\textbf{d}, Area-weighted mean absolute time-mean error and mean absolute biases in lag-one autocorrelation (AC1), temporal variance, skewness, and kurtosis for three observing networks, using HadCRUT5 masks and error covariances from 1948--2025, 1900--1977, and 1850--1927 applied to the same reference sequence.
Time-mean errors are calculated at each grid cell over unobserved months, and temporal statistics from complete reconstructed and reference time series, each defined as reconstruction minus reference.
AC1 is calculated after removing a linear trend; variance, skewness and excess kurtosis are calculated from the undetrended monthly anomaly series.
GenCRest-HF statistics are calculated from the 10-member ensemble mean.
Ellipse Kriging denotes the spatially adaptive, anisotropic Kriging method used in DCENT-I~\cite{chan2026dcent}, represented by its posterior mean.
Colors follow the common scale shown on the right, with lighter shades indicating smaller values.
Time-mean errors are in ${}^\circ\mathrm{C}$ and variance biases in ${}^\circ\mathrm{C}^{2}$; the remaining metrics are dimensionless.
The corresponding spatial patterns and global mean temperature residuals are shown in Extended Data Fig.~\ref{fig:temp5_jra3q_highstat} and Supplementary Figs.~B9, B10, B11, and~B12.
}
\label{fig:jra3qtemp}
\end{figure}

Historical climate observations are unevenly distributed in space, change in coverage over time, and contain correlated measurement and sampling errors.
HadCRUT5 illustrates these challenges: land stations and ship tracks leave extensive, time-varying gaps across land, oceans and polar regions (Fig.~\ref{fig:jra3qtemp}a,b).
Observational uncertainties also vary in space and time, and include both spatially correlated and uncorrelated components~\cite{kennedy2019ensemble,morice2021updated}.
Reliable historical reconstruction must therefore infer missing climate information while distinguishing climatic variability from observational noise.

To evaluate reconstruction under these conditions, we sample complete temperature fields using monthly HadCRUT5 masks and perturb the retained values with errors drawn from the corresponding observational-error covariance matrices~\cite{kennedy2019ensemble,morice2021updated}.
We use the Japanese Reanalysis for Three Quarters of a Century (JRA-3Q)~\cite{kosaka2024jra} as the reference, whose forecast model and data-assimilation system differ from those of ERA5, the reanalysis used to fine-tune GenCRest.
The complete, unperturbed fields allow reconstruction accuracy to be assessed under controlled observational coverage and uncertainty (see Methods and Supplementary Methods).
A representative month illustrates the differences among methods (Fig.~\ref{fig:jra3qtemp}c).
GenCRest reproduces coherent regional anomaly patterns and amplitudes close to the reference.
LaMa retains observational noise, consistent with treating observed values as exact during inpainting~\cite{kadow2020artificial,bochow2025reconstructing}.
Kriging suppresses noise but smooths the anomaly structure.

To assess whether our reconstruction preserves climate dynamics, we evaluate grid-cell time-mean errors alongside biases in lag-one autocorrelation (AC1), variance, skewness, and kurtosis.
AC1 and variance describe temporal persistence and fluctuation amplitude, whereas skewness and kurtosis characterize distributional asymmetry and tail behavior.
Preserving these properties is essential for investigating climate variability and extremes, and for analyses that use temporal statistics as indicators of Earth system resilience~\cite{scheffer2009early,boers2022theoretical,boers2025destabilization,liu2026SA}.

We reconstruct the same JRA-3Q temperature sequence over 1948--2025 using HadCRUT5 masks and error covariances from 1948--2025, 1900--1977, and 1850--1927, varying observational conditions while holding the reference climate evolution fixed (Fig.~\ref{fig:jra3qtemp}d; Supplementary Section~B.3.1).
GenCRest-HF yields the smallest mean absolute error and mean absolute bias over time across all higher-order statistics under the HadCRUT5 observing networks from all three historical periods (1948--2025, 1900--1977, and 1850--1927).
Its advantages in preserving temporal dynamics are particularly evident in higher-order statistics under the earliest observing network, when coverage was poorest.
LaMa underestimates temporal persistence (AC1) most strongly in the tropics, where persistent, small-amplitude climate signals are particularly susceptible to observational noise (Extended Data Fig.~\ref{fig:temp5_jra3q_highstat}b; Supplementary Figs.~B9b and~B10b).
Kriging underestimates the magnitude of temperature fluctuations (variance) over time, most strongly across the Arctic and Antarctic, where observational coverage is highly sparse (Extended Data Fig.~\ref{fig:temp5_jra3q_highstat}c; Supplementary Figs.~B9c and~B10c).
Spatially adaptive ellipse Kriging~\cite{chan2026dcent} reduces this underestimation, but its temporal-statistic biases remain larger than those of GenCRest-HF (Supplementary Fig.~B11).
For global mean temperature (GMT), GenCRest-LE generally outperforms the benchmarks at both monthly and annual timescales under the sparsest observing network (Supplementary Figs.~B10 and~B12).

These advantages of GenCRest persist across reference datasets.
Repeating the experiments with the 20CRv3 reanalysis, a CMCC-ESM2 simulation excluded from training, and ten held-out CMIP6 realizations yields consistent improvements in temporal statistics of climate dynamics, together with accurate GMT estimation (Supplementary Section~B.3.2; Supplementary Figs.~B15 and~B16).

GenCRest also preserves temporal dynamics and improves global land-mean precipitation (LMP) estimates in the more challenging reconstruction of high-resolution, non-Gaussian precipitation fields (Supplementary Figs.~B13 and~B14).
By contrast, ADW strongly suppresses temporal variance and reverts toward climatology in unobserved regions.

Together, these experiments show that GenCRest recovers both the local temporal statistics and the large-scale means of the reference climate from observations with realistic coverage and uncertainty, across variables, observing networks, and reference datasets.

\FloatBarrier
\section{Reconstruction of historical climate variability and extremes}\label{sec4}

We apply GenCRest to four observational records that underpin climate monitoring and assessment: HadCRUT5 and Berkeley Earth for temperature, and CRU TS 4.10 and GPCC v2025 for precipitation (Supplementary Methods).
The controlled experiments above demonstrate that GenCRest preserves temporal dynamics and spatial structure under sparse and uncertain observations, providing a more reliable basis for investigating regional variability and extremes that conventional infilling obscures.
The reconstructions are compared with the infilled products released by the data providers, hereafter the ``official infilled'' products.
They follow the common long-term evolution of GMT and LMP across products (Extended Data Fig.~\ref{fig:global_mean}), but reveal stronger regional variability and more extensive extremes during the sparsely observed early record, where spatial interpolation and climatological infilling in the official infilled products compress fluctuation amplitudes or assign near-zero anomalies.

\subsection{Temperature variability and extremes}

\begin{figure}[!htbp]
\centering
\includegraphics[width=\linewidth]{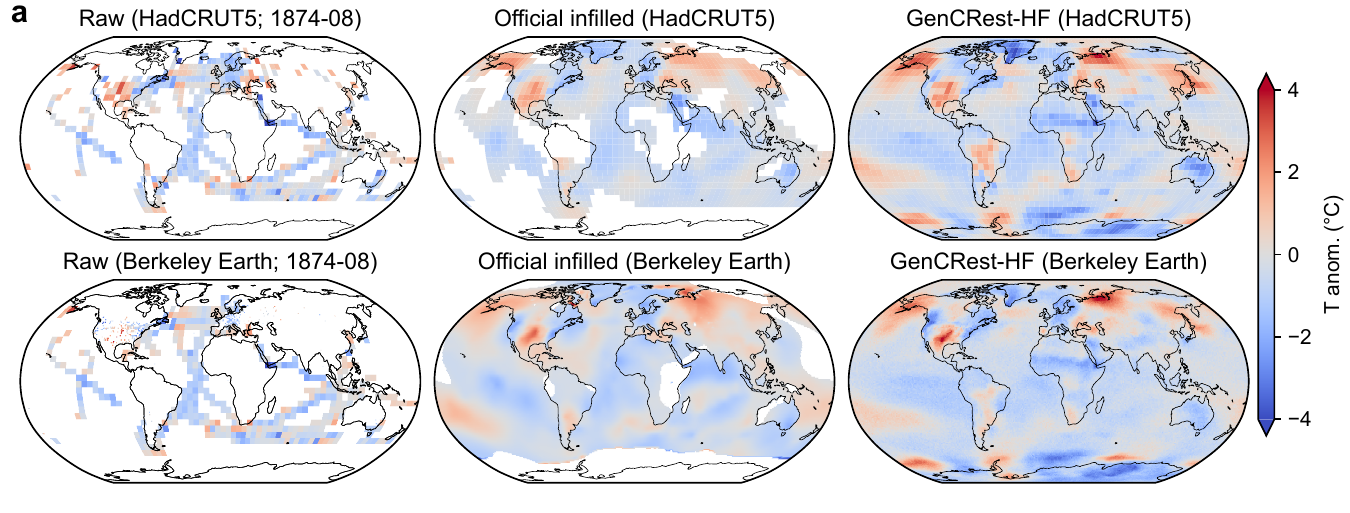}
\includegraphics[width=\linewidth]{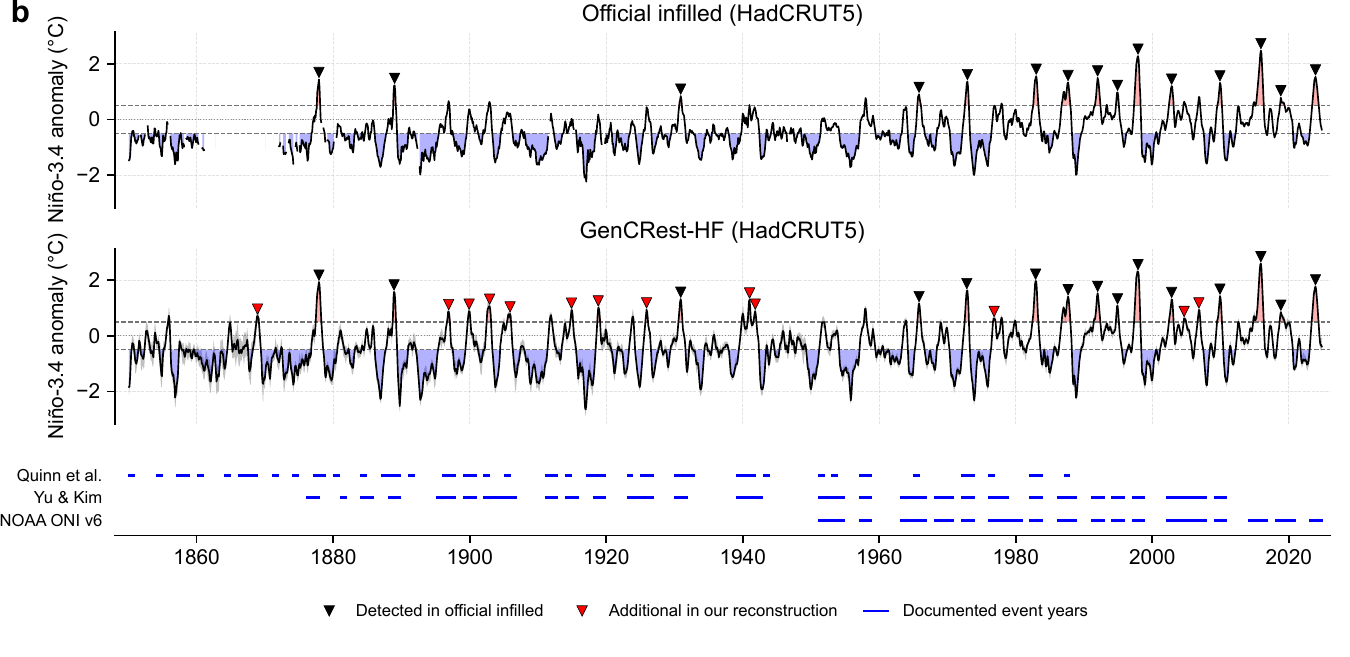}
\includegraphics[width=\linewidth]{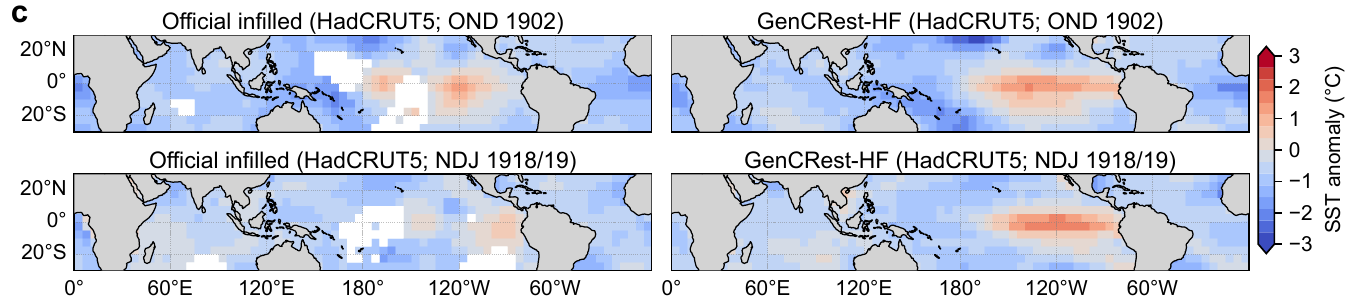}
\caption{
\textbf{Reconstruction of historical temperature patterns and El Ni\~no variability.}
\textbf{a}, August 1874 temperature anomalies from HadCRUT5 (top) and Berkeley Earth (bottom), relative to 1961--1990.
Columns show sparse observations (left), official infilled products (middle) and GenCRest-HF ensemble means (right).
The Berkeley Earth reconstruction combines station observations over land with HadSST4.2 observations~\cite{kennedy2019ensemble} over the ocean.
Additional historical examples and ensemble spread are shown in Supplementary Figs.~B18, B20 and~B21.
\textbf{b}, Ni\~no~3.4 indices for the HadCRUT5 official infilled product (top) and GenCRest-HF reconstruction (bottom) over 1850--2024, smoothed with a three-month running mean.
The reconstructed index is averaged across 10 members, with gray shading spanning their minimum and maximum.
Red and blue shading highlights values above $0.5\,{}^\circ\mathrm{C}$ and below $-0.5\,{}^\circ\mathrm{C}$, respectively.
El Ni\~no events require an index of at least $0.5\,{}^\circ\mathrm{C}$ for at least five consecutive overlapping three-month seasons.
Triangles mark peak seasons: black indicates events detected in the official product and temporally overlapping reconstructed events, whereas red indicates additional reconstructed events.
Blue lines below the indices show reference event years from the independent historical compilation of Quinn et al.~\cite{quinn1987nino} (confidence ratings $\geq 4$) and the SST-based records of Yu and Kim~\cite{yu2012identifying} and NOAA ONI v6~\cite{NOAA_ONI}.
Reference events are shown irrespective of detection in either series; blue horizontal bars indicate calendar-year ranges rather than precise event durations.
\textbf{c}, Three-month mean sea-surface temperature anomalies during October--December 1902 (top) and November 1918--January 1919 (bottom), comparing the HadCRUT5 official infilled product (left) with the 10-member GenCRest-HF ensemble mean (right).
Anomalies in \textbf{b} and \textbf{c} are relative to 1991--2020.
The corresponding Berkeley Earth analysis is shown in Extended Data Fig.~\ref{fig:berkeley_enso}.
}
\label{fig:temp_recon}
\end{figure}

The temperature reconstructions reveal larger-amplitude anomalies and sharper regional contrasts than the official infilled products.
In August 1874, for example, pronounced anomalies extend across the Arctic, Eurasia, and the Southern Ocean, with additional spatial detail in the $1^\circ$ Berkeley Earth reconstruction (Fig.~\ref{fig:temp_recon}a).
These contrasts recur across historical examples and diminish as observational coverage expands (Supplementary Section~B.4.1).
The greater regional variability is accompanied by more extensive hot extremes during much of 1850--1950, consistent with controlled experiments in which GenCRest-HF preserves spatial variability and hot-extreme coverage under historical observing conditions (Supplementary Figs.~B22, B23, and~B24; Supplementary Section~B.4.2).

In the tropical Pacific, we identify stronger early-twentieth-century ENSO variability, with larger warm-phase Ni\~no~3.4 anomalies in both the HadCRUT5 and Berkeley Earth reconstructions (Fig.~\ref{fig:temp_recon}b; Extended Data Fig.~\ref{fig:berkeley_enso}a,b).
Over 1850--2024, our reconstructions identify approximately 50\% more historically documented El Ni\~no events than the official infilled products.
Episodes such as 1902 and 1918--1919~\cite{quinn1987nino,yu2012identifying,giese20101918} exhibit coherent warming across the central and eastern equatorial Pacific, where the official fields show weaker and less continuous anomalies (Fig.~\ref{fig:temp_recon}c; Extended Data Fig.~\ref{fig:berkeley_enso}c).
The greater variability of our reconstructions is supported by station-based atmospheric records, which are independent of marine observations. 
The reconstructed Ni\~no~3.4 anomalies correlate more strongly with the Southern Oscillation Index (SOI) and with Darwin and Tahiti pressure than the official infilled products do, most markedly before 1920, when SOI-identified El Ni\~no episodes are frequent and coincide with the reconstructed warm phases (Supplementary Fig.~B29; Supplementary Table~B6).
The same patterns recur with alternative sea surface temperature (SST) inputs and are consistent with controlled experiments showing that conventional interpolation attenuates El Ni\~no amplitudes (Supplementary Section~B.4.3).

\subsection{Precipitation variability and extremes}

\begin{figure}[!htbp]
\centering
\includegraphics[width=1.\linewidth]{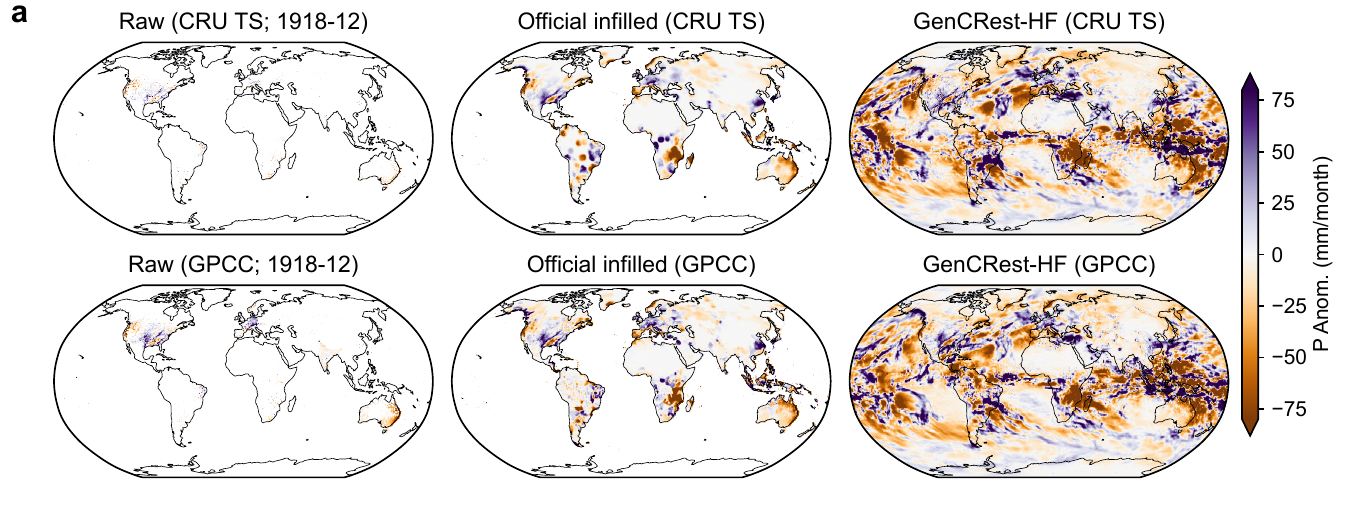}
\includegraphics[width=1.\linewidth]{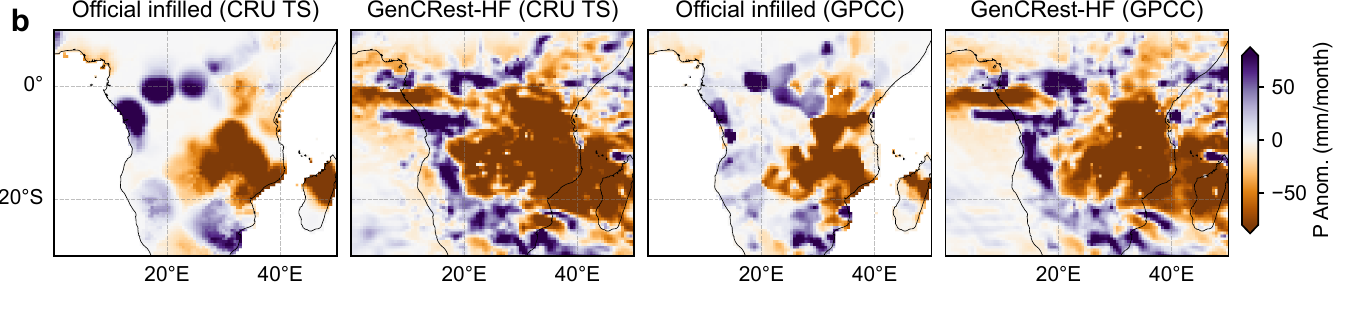}
\includegraphics[width=1.\linewidth]{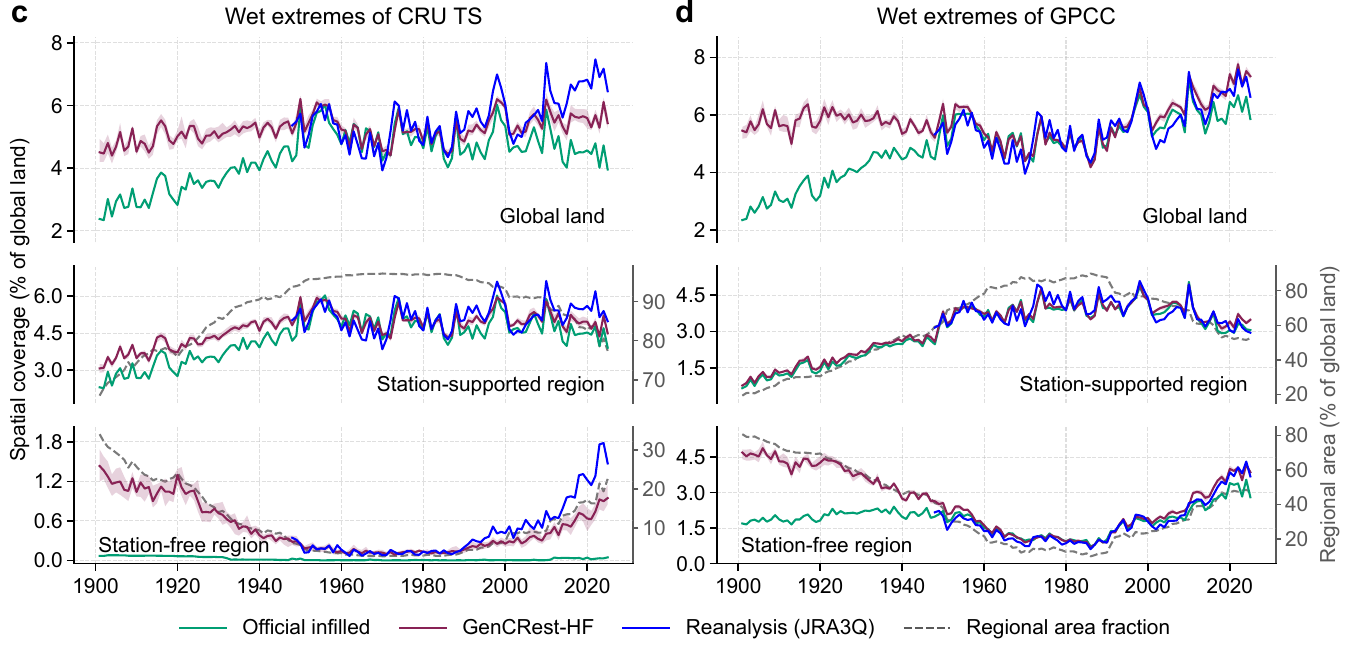}
\caption{
\textbf{Reconstruction of historical precipitation patterns and wet extremes.}
\textbf{a}, December 1918 precipitation anomalies relative to 1961--1990.
Rows show CRU TS 4.10 and GPCC v2025; columns show sparse conditioning fields, official infilled products and representative single GenCRest-HF members.
The GPCC conditioning field consists of gauge-supported gridded estimates rather than individual gauge measurements.
Ocean anomalies are inferred from land-only conditioning.
Additional historical examples and ensemble spread are shown in Supplementary Figs.~B32, B34 and~B35.
\textbf{b}, Corresponding fields over southern Africa.
\textbf{c}, \textbf{d}, Annual wet-extreme coverage for CRU TS 4.10 (\textbf{c}) and GPCC v2025 (\textbf{d}).
Wet extremes are defined separately for each product and reconstruction member as monthly precipitation anomalies exceeding the local 95th percentile for 1961--1990.
Annual coverage is the mean monthly fraction of the corresponding product's valid land area experiencing wet extremes.
Rows show total global-land wet-extreme coverage and contributions from station-supported and station-free regions, defined using product-specific criteria.
For CRU TS, the two classes correspond to positive and zero station-support counts, respectively.
For GPCC, unchanged and changed precipitation estimates following climatological infilling are used as proxies for the two support classes, respectively.
Dashed gray curves show the area fractions of the two classified regions, normalized by their combined area.
Purple curves are an average of the coverages calculated separately from 10 GenCRest-HF members; green and blue curves show the official infilled products and JRA-3Q, respectively.
Shading spans the minimum and maximum of the member-specific annual coverages.
Station-support definitions, threshold construction and aggregation are provided in Supplementary Section~B.5.3.
}
\label{fig:precip_recon}
\end{figure}

Our precipitation reconstructions reveal pronounced local wet and dry anomalies over sparsely observed land regions.
In December 1918, for example, such features are evident over southern Africa, and they recur across South America, China, and Siberia, where the official infilled products show smoother anomalies or radial patterns (Fig.~\ref{fig:precip_recon}a,b; Supplementary Figs.~B32 and~B33).

Wet extremes generally cover a larger fraction of global land in our reconstructions than in the official infilled products, with the largest differences in the early twentieth century, when station-free regions were most extensive (Fig.~\ref{fig:precip_recon}c,d).
The difference is concentrated in CRU TS regions without station support and GPCC regions affected by climatological infilling (the bottom row in Fig.~\ref{fig:precip_recon}c,d), whereas the estimates evolve similarly where observational support is stronger.
Climatological replacement suppresses anomaly variability in these regions, reducing their contribution to the inferred extent of wet extremes.
Removing climatological infilling from the alternative GPCC analysis brings wet-extreme coverage closer to our reconstructions over much of the record, and the reconstructed coverage follows reanalysis more closely during their overlap (Supplementary Section~B.5.3).

Conditioning on land observations alone also provides useful estimates of precipitation variability over the global ocean, extending precipitation coverage beyond the land-only official precipitation products to globally complete fields.
During the satellite era (1980--2025), the inferred ocean fields show qualitative agreement with the broad wet and dry anomaly patterns of satellite and reanalysis estimates, together with similarly shaped normalized spatial spectra (Extended Data Fig.~\ref{fig:precip_gpcp_jra3q}; Supplementary Fig.~B36).
Ensemble spread over the oceans further identifies regions where these estimates remain uncertain (Supplementary Figs.~B34 and~B35).
This agreement between land-conditioned reconstructions and ocean precipitation patterns is consistent with learned long-range dependencies, potentially including global precipitation teleconnections~\cite{boers2019complex}.

\section{Reassessment of long-term climate change}\label{sec5}

The recovered regional variability raises the question of whether long-term trends, the basis of climate-change assessments, are likewise affected by how observational gaps are filled.
We therefore examine temperature trends over 1900--1980 and 1981--2024 and precipitation trends over 1901--2025 and 1980--2025, following the temporal framing of IPCC AR6~\cite{gulev2021changing}, and compare them with the official infilled products and with complementary satellite, reanalysis, and station records.

\subsection{Temperature trends and Arctic warming}

\begin{figure}[htbp]
\centering
\includegraphics[width=.98\linewidth]{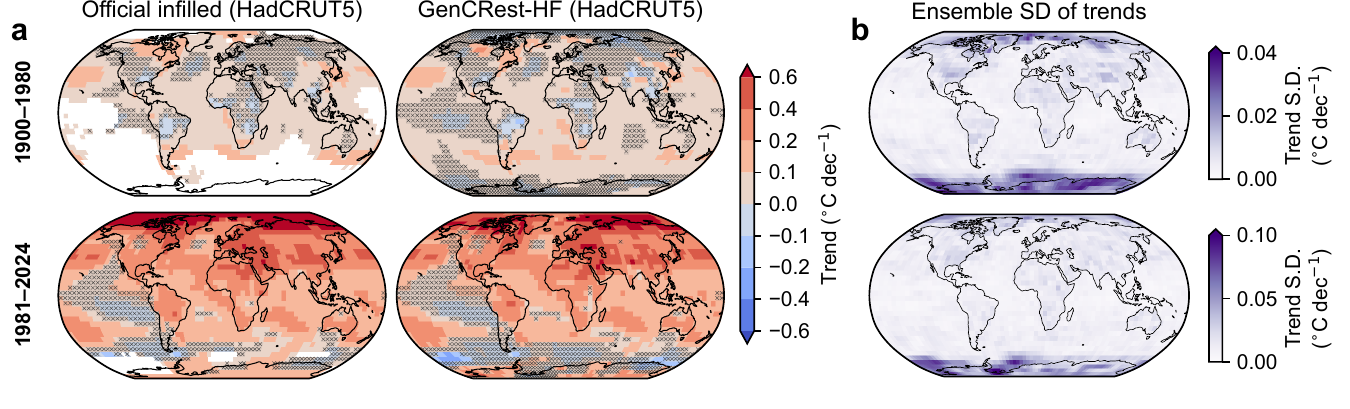}
\includegraphics[width=.49\linewidth]{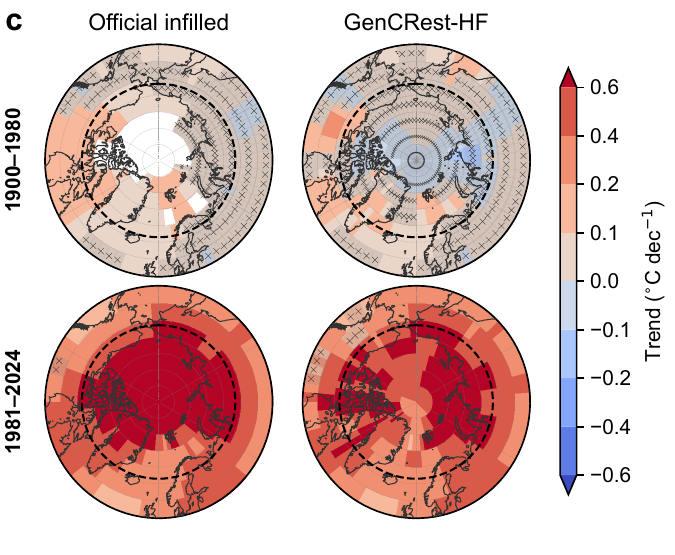}
\includegraphics[width=.49\linewidth]{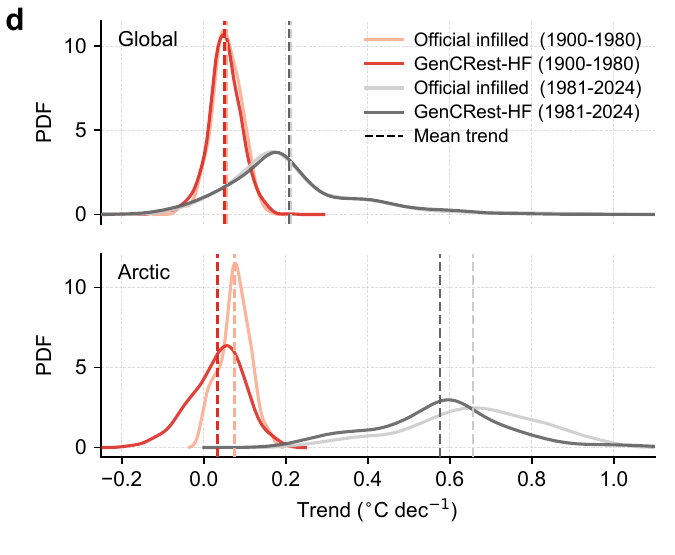}
\includegraphics[width=.98\linewidth]{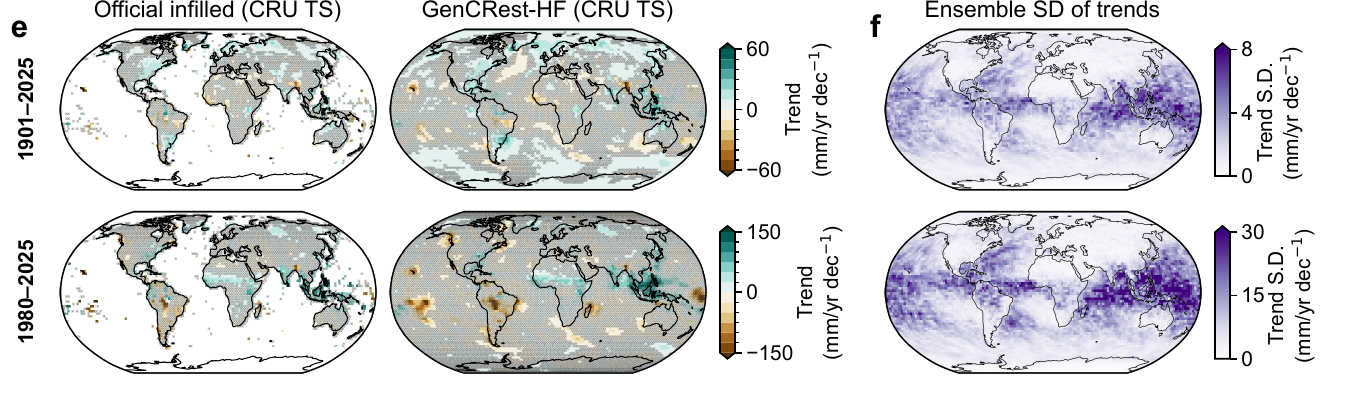}
\includegraphics[width=.98\linewidth]{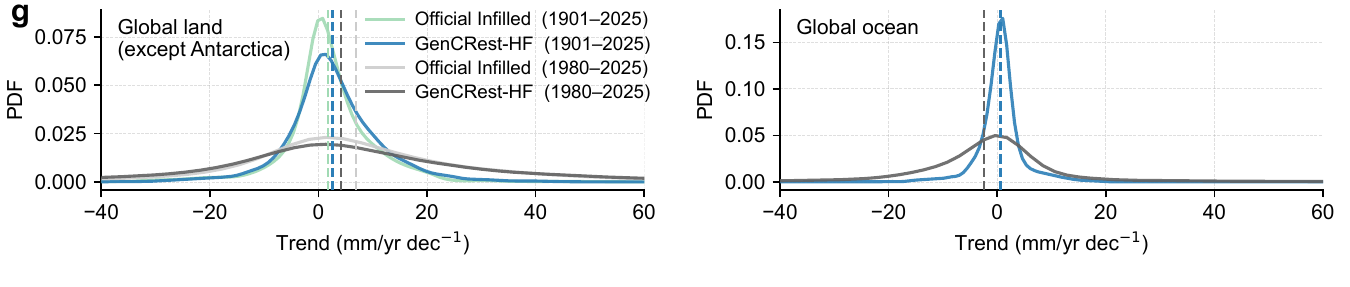}
\caption{
\textbf{Spatial patterns and distributions of historical temperature and precipitation trends.}
HadCRUT5 temperature trends are evaluated on the native $5^\circ$ grid, and CRU TS 4.10 precipitation fields are aggregated to $2.5^\circ$ before calculating trends and ensemble spread.
\textbf{a}, Linear temperature trends in the HadCRUT5 official infilled product and GenCRest-HF reconstruction during 1900--1980 (top) and 1981--2024 (bottom).
Reconstructed trends are calculated from the 10-member ensemble-mean field.
\textbf{b}, Ensemble spread, expressed as the standard deviation of temperature trends calculated separately for each reconstruction member.
\textbf{c}, Corresponding Arctic temperature trends, with dashed circles marking $66.5^\circ$\,N.
\textbf{d}, Area-weighted probability density functions (PDFs) of temperature trends across the globe (top) and the Arctic (bottom).
In \textbf{d}, the reconstruction is restricted to the valid spatial coverage of the official infilled product for each period, so both temperature distributions represent the same grid cells.
\textbf{e}, Linear trends in annual precipitation from the CRU TS 4.10 official infilled product and GenCRest-HF reconstruction during 1901--2025 (top) and 1980--2025 (bottom).
Reconstructed trends are averaged across 10 GenCRest-HF members.
\textbf{f}, Corresponding ensemble spread in precipitation trends, calculated as in \textbf{b}.
\textbf{g}, Area-weighted PDFs of precipitation trends over global land excluding Antarctica (left) and of reconstructed trends over the global ocean (right).
The distributions in \textbf{d} and \textbf{g} include all valid trends, with vertical dashed lines indicating their area-weighted means.
Each precipitation distribution in \textbf{g} uses the valid grid cells available for that field within the stated land or ocean domain.
Gray crosses in \textbf{a}, \textbf{c} and \textbf{e} mark trends that do not meet the 5\% false-discovery-rate criterion after adjustment for temporal autocorrelation~\cite{santer2008consistency,wilks2016stippling}.
}
\label{fig:largescale_trend}
\end{figure}

Our reconstructions indicate a more spatially heterogeneous history of temperature change than the official infilled products, particularly in the Arctic.
During the sparsely observed period 1900--1980, both the HadCRUT5 and Berkeley Earth reconstructions show pronounced regional warming and cooling, including warming around Greenland and the Barents Sea that is supported by other independent gridded products~\cite{yin2024noaaglobaltemp,lenssen2024nasa,calvert2024improving,rohde2026besthr} (Fig.~\ref{fig:largescale_trend}a--d; Extended Data Fig.~\ref{fig:berkeley_gpcc_largescale_trend}a--d; Supplementary Fig.~B41).
Ensemble spread identifies the regions where these early trends are less tightly constrained by observations (Supplementary Fig.~B39).

During 1981--2024, our reconstructions reproduce the global trend distributions of the official infilled products but yield lower Arctic-mean warming, by 0.08 and 0.12~${}^\circ\mathrm{C}\,\mathrm{decade}^{-1}$ for HadCRUT5 and Berkeley Earth, respectively (Fig.~\ref{fig:largescale_trend}d; Extended Data Fig.~\ref{fig:berkeley_gpcc_largescale_trend}d).
The official infilled products show spatially uniform warming across the central Arctic, whereas the reconstructions show weaker warming over parts of the interior and localized maxima over Greenland and the Barents and Kara Seas (Fig.~\ref{fig:largescale_trend}c; Extended Data Fig.~\ref{fig:berkeley_gpcc_largescale_trend}c).
Arctic amplification, the ratio of Arctic to global warming~\cite{rantanen2022arctic}, shifts accordingly: for HadCRUT5, the pan-Arctic factor falls from 3.12 in the official product to 2.84, while median amplification across grid cells rises from 2.70 to 3.38 in the Barents Sea and from 4.08 to 4.39 in the Kara Sea (Extended Data Fig.~\ref{fig:temp5_arctic_AA}).
This contrast may reflect spatial smoothing during statistical interpolation.
Such smoothing can weaken localized warming maxima while extending warming signals from better-observed surrounding regions into the poorly observed central Arctic.
Lower Arctic-mean warming and pronounced regional contrasts persist with CMIP6-only and ERA5-only priors, as well as when post-1980 fields are excluded from training and when air and sea-surface temperature are modeled separately (Supplementary Fig.~B46; Supplementary Sections~B.6.4 and~B.6.5).

The reconstructed pattern of Arctic warming is corroborated by complementary references.
Both reconstructions agree more closely than their official counterparts with reanalysis near-surface temperature, satellite skin temperature, and stations across the Arctic (Extended Data Fig.~\ref{fig:temp5_arctic_reference}; Supplementary Section~B.6.2; Supplementary Figs.~B43 and~B44).
Most other gridded instrumental products share the near-uniform central-Arctic warming of the official infilled products and agree less well with these references (Supplementary Fig.~B41).
Only two recent products reach comparable agreement, through temperature-specific designs that import spatial patterns from reanalysis or explicitly represent spatially varying warming and sea-ice effects~\cite{calvert2024improving,rohde2026besthr}, indicating that a general-purpose generative reconstruction attains similar regional fidelity without variable-specific design.

\subsection{Precipitation trends across land and ocean}

Our reconstructions indicate stronger regional contrasts in long-term precipitation change than the official infilled products, over land and ocean alike.
Over 1901--2025, both reconstructions yield broader land-trend distributions with less mass near zero, indicating more pronounced regional wetting and drying (Fig.~\ref{fig:largescale_trend}e--g; Extended Data Fig.~\ref{fig:berkeley_gpcc_largescale_trend}e--g).
Over 1980--2025, the land-trend distributions of our reconstructions and the official infilled products converge, but across both land and ocean, the reconstructed distributions are broader than over the full record.
Regional trends have thus intensified in recent decades, a change that is masked in land- and ocean-wide averages, where wetting and drying offset one another (Fig.~\ref{fig:largescale_trend}g; Extended Data Fig.~\ref{fig:berkeley_gpcc_largescale_trend}g).

These regional trends are supported by independent reanalysis and satellite records.
Over land, both reconstructions agree more closely with the reanalysis and satellite than the official infilled products do, particularly in under-monitored regions (Supplementary Section~B.7; Supplementary Fig.~B51).
Over the ocean, where the reconstructions are informed only by land observations, trend amplitudes are compressed toward zero, but the spatial pattern of change is moderately correlated with the satellite record (Extended Data Fig.~\ref{fig:ocean_precip_scatter}).
Agreement is highest at intermediate signal-to-noise ratios and declines in the highest signal-to-noise ratio bin, a dependence shared by the reanalysis despite its full observational assimilation.

\FloatBarrier
\section{Discussion}

Our results show that GenCRest can reconstruct the spatial structure and temporal variability of global temperature and precipitation fields from sparse and uncertain observations.
For temperature, conventional interpolation smooths climatic fluctuations where observations are sparse~\cite{yamamoto2005correcting,rohde2013new,harris2020version}, whereas deterministic inpainting retains observational noise when measured values are treated as exact~\cite{kadow2020artificial,bochow2025reconstructing}.
For precipitation, existing approaches have great difficulty in producing realistic spatial patterns. 
GenCRest combines a learned spatiotemporal prior with observation guidance that accounts for measurement uncertainty.
The generative modeling helps preserve fluctuation amplitudes and non-Gaussian variability, while temporal context further improves spatial accuracy, temporal persistence, and higher-order statistics (Supplementary Section~B.3.4).
These complementary contributions improve the representation of climate variability and extremes under changing observational coverage.

In GenCRest, the learned prior and the observations play distinct roles.
The prior supplies statistical relationships of climate states across space and time, through which observations in one region inform the reconstruction elsewhere.
Pre-training on a multimodel CMIP6 archive exposes GenCRest to diverse climate histories, and fine-tuning on ERA5 attenuates potential effects of model bias and adapts the learned distribution to an observationally constrained representation of the climate system, with complementary benefits for reconstruction accuracy and temporal statistics (Supplementary Section~B.3.5).
Because training uses randomly sampled short segments without calendar-time information, GenCRest is not explicitly trained to map historical dates to long-term climate trends.
Robustness assessments using alternative training sources and periods support the robustness of our findings (Supplementary Section~B.6.4).
Nevertheless, the observation-constrained reconstructions are not fully independent of the climate models used to train the prior.
This dependence should be considered when using the reconstructions for model--data comparisons, particularly in regions and periods with sparse observations.

Historical evolution, in turn, enters through the observations used to condition the learned prior.
GenCRest incorporates observational uncertainty through the error covariance matrix in the likelihood, allowing departures from uncertain measurements and accounting for error correlations where available.
Our downstream reconstruction builds on upstream data collection, quality control, homogenization, and uncertainty estimation by observational-data providers~\cite{morice2021updated,rohde2020berkeley,kennedy2019ensemble,harris2020version,schneider2014gpcc}.
Closer collaboration among observational data specialists, climate scientists, and machine learning researchers can refine observational corrections and uncertainty estimates, which can be incorporated directly into conditional sampling without retraining the generative prior.

Joint reconstruction of temperature, precipitation, circulation, and sea ice, with physical constraints, could improve consistency among variables and support studies of the mechanisms underlying regional variability of extremes.
Diffusion sampling remains computationally demanding~\cite{li2024learning}, requiring choices about how much computation to devote to each realization and how many ensemble members to generate.
GenCRest allows the sampling effort per realization and the ensemble size to be adjusted according to available computational resources and scientific objectives.
We illustrate two reference configurations (i.e., GenCRest-HF and GenCRest-LE) for allocating this effort within the same trained model.
More efficient diffusion solvers, latent-space diffusion, and model distillation could reduce the cost of generating large ensembles while preserving the spatial structure and temporal variability of individual realizations.

Finally, by preserving spatiotemporal consistency where observations are sparse, our reconstructions reveal climate variability and extremes that conventional infilling attenuates, as illustrated by stronger early-twentieth-century ENSO variability, more widespread historical wet extremes, and more heterogeneous Arctic warming.
These signals broaden the historical baseline against which recent climate anomalies are assessed, while the accompanying ensembles allow explicit examination of reconstruction uncertainty.
Extending GenCRest to other sparsely observed climate variables could reveal previously unresolved historical changes and help identify the mechanisms underlying regional climate variability and extremes.

\clearpage
\begin{figure}[!htbp]
\centering
\includegraphics[width=\linewidth]{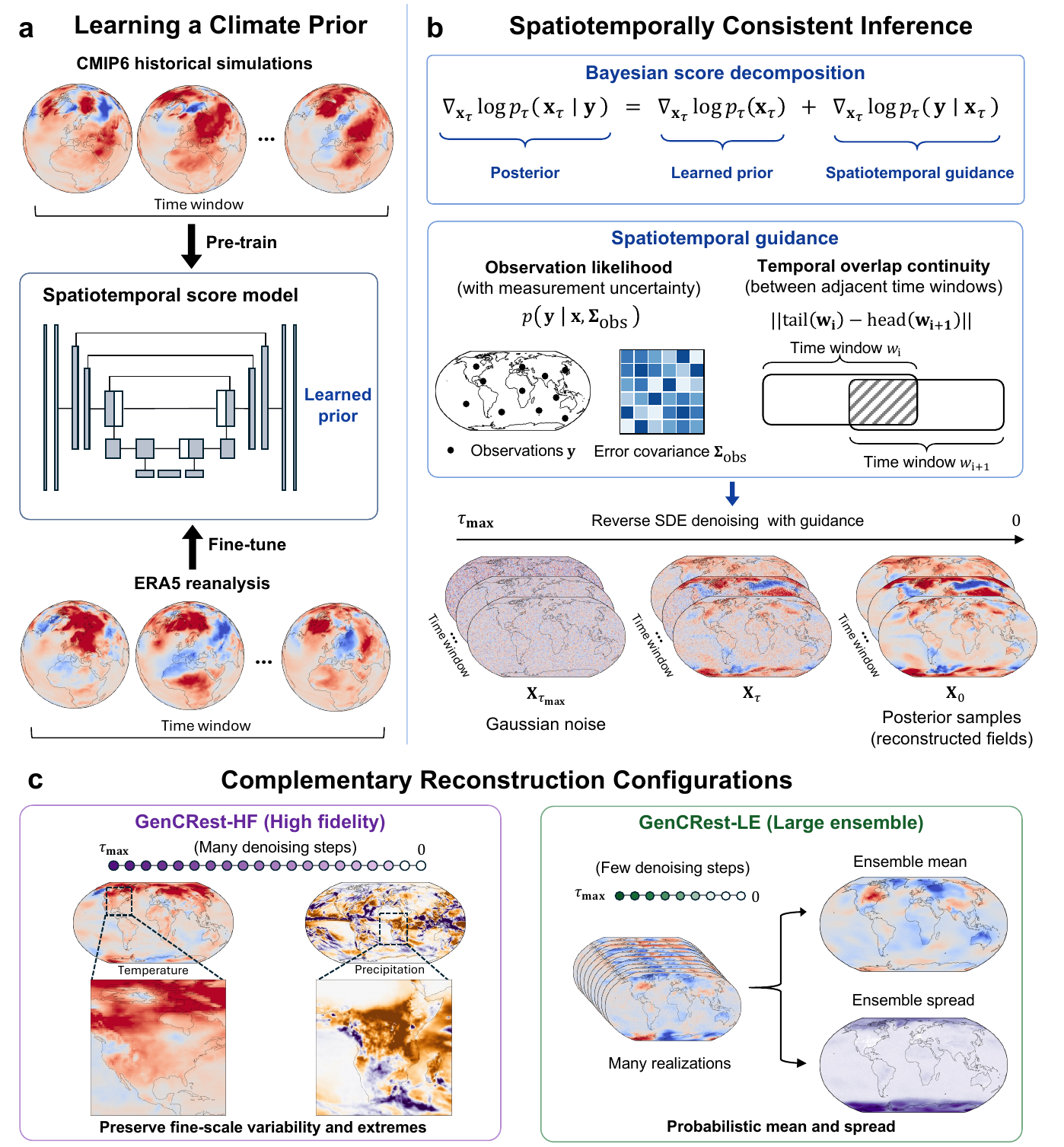}
\caption{
\textbf{GenCRest for probabilistic, spatiotemporally consistent climate reconstruction.}
\textbf{a}, Learning a spatiotemporal climate prior.
A score model is pre-trained on continuous time windows from CMIP6 historical simulations and fine-tuned on ERA5 reanalysis to represent the joint spatial and temporal variability of complete temperature and precipitation fields.
\textbf{b}, Observation-conditioned reconstruction.
Starting from Gaussian noise, reverse-time stochastic differential equation (SDE) sampling combines the learned prior with observation guidance and temporal-overlap guidance.
The observation likelihood incorporates measurement uncertainty through the covariance matrix $\boldsymbol{\Sigma}_{\mathrm{obs}}$.
The temporal term reduces discrepancies between overlapping portions of adjacent reconstruction windows.
Repeated diffusion denoising steps generate an ensemble of complete fields conditioned on the observations.
\textbf{c}, Two conditional sampling configurations of GenCRest.
Within each variable-resolution setting, GenCRest-HF and GenCRest-LE share the same trained model and observations but use different denoising-step counts and guidance weights.
Under comparable total computational budgets, GenCRest-HF generates fewer realizations with more denoising steps per realization, whereas GenCRest-LE generates more realizations with fewer steps each.
Both configurations produce complete spatiotemporal realizations that can be analyzed individually or summarized by ensemble means and spread.
}
\label{fig:flowchart}
\end{figure}

\clearpage

\clearpage

\clearpage

\clearpage

\clearpage

\clearpage
\section{Methods}\label{Method}
\subsection{Probabilistic generative reconstruction}\label{method_reconintro}

Our objective is to reconstruct complete historical climate fields from sparse, fragmented observations.
We frame reconstruction as a probabilistic inverse problem: estimating the conditional distribution $p(\mathbf{x} \mid \mathbf{y})$ of the full climate state $\mathbf{x}$ given sparse observations $\mathbf{y}$, considering the intrinsic chaos and multiscale variability of the climate system.
Using this formulation, GenCRest (Fig.~\ref{fig:flowchart}) generates posterior ensembles that quantify uncertainty and preserve the spatiotemporal and higher-order statistical characteristics of climate variability.

\textbf{Generative climate prior.}
GenCRest uses a video diffusion model~\cite{song2020score,ho2022video} to learn a prior distribution over high-dimensional spatiotemporal climate fields.
It combines this learned prior with observational likelihood through conditional sampling to address the corresponding Bayesian inverse problem (see Supplementary Section~A.1 for details).
Adopting a spatiotemporal U-Net architecture, the diffusion model approximates the score function $\nabla_{\mathbf{x}} \log p(\mathbf{x})$ to learn the spatiotemporal joint distribution of climate variability.
We first pre-train the model using historical CMIP6 simulations, then fine-tune it with ERA5 reanalysis, both performed unconditionally.
This training strategy allows the model to learn the complex dynamics of the Earth system, independent of observation locations.

\textbf{Conditional inference strategy.}
During reconstruction, GenCRest generates complete climate fields by numerically integrating the reverse-time stochastic differential equation (SDE) from Gaussian noise.
Observation guidance incorporates the available measurements, while temporal guidance promotes continuity between overlapping reconstruction segments (see Supplementary Section~A.2 for details).

\textbf{Conditional sampling configurations.}
Likelihood-guided posterior sampling progressively refines each realization through denoising and observational and temporal guidance~\cite{li2024learning}.
Increasing the number of denoising steps can improve the accuracy of individual realizations, at greater computational cost.
The sampling effort per realization and ensemble size can therefore be adjusted to suit scientific priorities and available computational resources.
Here, we illustrate this flexibility through GenCRest-HF and GenCRest-LE, two sampling configurations that allocate comparable total computational budgets differently between individual refinement and ensemble size.
Within each variable-resolution setting, both configurations share the same trained GenCRest model, training objective, and observational inputs, but use separately selected sampling-step counts and guidance weights.

GenCRest-HF uses more denoising steps per realization to generate a few ensemble members.
The additional refinement improves the representation of local spatial structure, temporal evolution, and higher-order variability, supporting studies of regional climate processes and extremes.
GenCRest-LE uses fewer steps per realization to generate many members, providing a larger ensemble for estimating posterior expectations and reconstruction uncertainty.
Its individual realizations retain regional structure and temporal variability, although these features are generally represented less faithfully than in the longer GenCRest-HF sampling.
We use GenCRest-LE primarily to examine mean climate states (e.g., global mean temperature) and their uncertainty.

Both configurations support analyses of individual realizations and ensemble statistics.
We use 10 members for GenCRest-HF and 50 for GenCRest-LE, with matched total numbers of denoising steps within each reconstruction setting.
An ensemble-size sensitivity analysis shows rapid initial improvements in ensemble-mean accuracy followed by diminishing gains as more members are included (Supplementary Fig.~A1).
These settings illustrate practical sampling choices: longer sampling trajectories can be combined with larger ensembles when computational resources permit or when a specific scientific purpose requires it.
The full sampling settings are reported in Supplementary Table~A1.

\subsection{Training data and preprocessing}
\label{method_trainingdata}

GenCRest is trained sequentially using historical CMIP6 simulations and ERA5 reanalysis.
The training datasets contain monthly mean surface air temperature (\textit{tas} and \textit{t2m}) and precipitation (\textit{pr} and \textit{tp}) fields.

For the initial pre-training, we use 20 historical simulations from ten CMIP6 models over 1850--2014.
These comprise the \textit{r1i1p1f1} and \textit{r2i1p1f1} realizations from each of ACCESS-CM2, BCC-CSM2-MR, CESM2, EC-Earth3, FGOALS-f3-L, GFDL-ESM4, MIROC6, MPI-ESM1-2-HR, MRI-ESM2-0 and NorESM2-MM.
The multimodel archive broadens the range of climate trajectories represented in the learned prior, including differences in spatial patterns, temporal variability and responses to external forcing.
By selecting the same number of realizations from each model, we also avoid disproportionately weighting models for which larger ensembles are available.

For temperature reconstruction, GenCRest is fine-tuned on ERA5 reanalysis after CMIP6 pre-training.
This stage brings the learned spatial covariance, temporal variability, and evolution statistics closer to those represented in an observationally constrained reanalysis.
For precipitation reconstruction, GenCRest undergoes an additional intermediate training stage.
The model is first pre-trained on the same 20 standard-resolution CMIP6 simulations, which provide a broad, large-scale hydroclimatic prior.
It is then further trained on four higher-resolution precipitation simulations over 1950--2014: CNRM-CM6-1-HR \textit{r1i1p1f2}, ECMWF-IFS-HR \textit{r1i1p1f1}, GFDL-CM4C192 \textit{r1i1p1f1} and HadGEM3-GC31-HM \textit{r1i1p1f1}.
The enhanced horizontal resolution provides a richer representation of storm-associated and convection-related precipitation structures than the standard-resolution CMIP6 simulations.
Finally, the model is fine-tuned on ERA5 precipitation to align the learned spatiotemporal statistics more closely with those represented in reanalysis.
The precipitation training sequence, therefore, consists of standard-resolution CMIP6 pre-training, higher-resolution intermediate training, and ERA5 fine-tuning.

All fields are remapped to regular latitude--longitude grids using conservative remapping.
Temperature fields are remapped to both $5^\circ$ ($36\times72$ pixels) and $1^\circ$ ($180\times360$ pixels) grids, whereas precipitation fields are remapped to a $0.5^\circ$ ($360\times720$ pixels) grid.
For each variable and spatial resolution, we compute monthly anomalies relative to the 1961--1990 climatological baseline.
The period 2005--2014 is strictly excluded from all training stages and reserved as the test set.

To reduce the strong skewness and heavy tails of precipitation anomalies while retaining both positive and negative values, we apply a scaled inverse hyperbolic sine transformation~\cite{burbidge1988alternative}:
\begin{equation}
\label{eq:precip_asinh}P^{*}=\operatorname{arcsinh}\left(\frac{P}{c_{\mathrm{pr}}}\right),\end{equation}
where $P$ denotes the precipitation anomaly and $c_{\mathrm{pr}}=0.5\sigma_{\mathrm{ERA5}}$, with $\sigma_{\mathrm{ERA5}}$ denoting the standard deviation of ERA5 precipitation anomalies in the training data.
The transformation is approximately linear for small anomalies but compresses the magnitudes of both positive and negative extremes.
Finally, all fields are standardized by subtracting the global mean and dividing by the global standard deviation.
These statistics are computed from the corresponding training data for each variable and spatial resolution and are held fixed during validation and testing.

\subsection{Benchmarking and evaluation}\label{benchmark}

In the idealized gap-filling experiments, we apply synthetic observation masks with random uniform coverage over space and time to the ERA5 and CMIP6 test data over 2005--2014, evaluating reconstruction across a range of observational coverages.
For temperature, the primary comparisons include Kriging~\cite{rohde2013new} and the deep-learning inpainting model LaMa~\cite{bochow2025reconstructing}.
We additionally compare the $5^\circ$ temperature reconstruction with a Fourier neural operator~\cite{li2020fourier}, a graph neural network~\cite{wu2021inductive}, DeepKriging~\cite{chen2024deepkriging} and ellipse Kriging~\cite{chan2026dcent}.
For precipitation, we compare against angular-distance weighting (ADW)~\cite{harris2020version} and LaMa.
All methods are evaluated at the spatial resolutions corresponding to the target reconstruction products: $5^\circ$ and $1^\circ$ for temperature and $0.5^\circ$ for precipitation.

We assess out-of-sample generalization using four groups of climate records.
These comprise the JRA-3Q reanalysis over 1948--2025, the 20CRv3 reanalysis over 1850--2015, an independent CMCC-ESM2 \textit{r1i1p1f1} simulation over 1850--2014, and the \textit{r3i1p1f1} simulations from each of the ten CMIP6 models represented during pre-training over 1850--2014.
JRA-3Q and 20CRv3 provide evaluations based on independent reanalysis systems, whereas CMCC-ESM2 evaluates generalization to a climate model that is entirely absent from the training archive.
The ten \textit{r3i1p1f1} simulations were not used for training and evaluate generalization to independent climate trajectories from model families represented during pre-training.

We assess reconstruction performance using metrics of temporal coherence and spatial accuracy.
Temporal correlations, spatial nRMSE and grid-cell time-mean errors are evaluated at unobserved grid--time points, whereas higher-order temporal statistics use complete reconstructed and reference time series.
Let $\mathbf{Y}\in\mathbb{R}^{T\times H\times W}$ denote the ground-truth field, $\hat{\mathbf{Y}}$ the reconstructed field and $\mathbf{M}$ the binary observation mask, with $\mathbf{M}_{t,i,j}=1$ for observed values and $\mathbf{M}_{t,i,j}=0$ for unobserved values.

\textbf{Temporal coherence.}
We quantify the temporal accuracy of the reconstruction using the mean temporal correlation coefficient (TCC).
We calculate the Pearson correlation coefficient ($r$) between the reconstructed and ground truth anomaly time series for each individual grid cell $(i, j)$ within the target region.
The global metric is reported as the latitude-weighted mean of these local correlations:

\begin{equation}
\text{Mean TCC} = \frac{1}{\sum_{(i,j) \in \mathcal{V}} w_{i,j}} \sum_{(i,j) \in \mathcal{V}} w_{i,j} \cdot \rho \left( \hat{\mathbf{Y}}_{:, i, j}[\Omega_{i,j}], \mathbf{Y}_{:, i, j}[\Omega_{i,j}] \right)
\end{equation}

\noindent where $\mathcal{V}$ is the set of valid grid cells (where correlation is defined), $w_{i,j} = \cos(\phi_i)$ denotes the latitude weight at grid cell $(i,j)$, $\phi_i$ is the corresponding latitude, $\Omega_{i,j}$ represents the set of time indices where the mask is zero for location $(i,j)$, and $\rho$ denotes the Pearson correlation function.
A higher mean TCC indicates that the model correctly captures the evolution of climate variability over time, with a theoretical maximum of 1.

\textbf{Spatial accuracy.}
We evaluate the reconstruction of spatial anomaly patterns using the spatial normalized root mean square error (nRMSE).
At each time step $t$, the latitude-weighted RMSE is calculated over the unobserved spatial domain $\mathcal{S}_t$ and normalized by the weighted spatial standard deviation of the ground-truth anomalies over the same domain.
The final score is averaged across all time steps:

\begin{equation}
\label{eq:mean_nrmse}
\mathrm{Mean\ nRMSE}=\frac{1}{T}\sum_{t=1}^{T}\frac{\sqrt{\frac{\sum_{(i,j)\in\mathcal{S}_t}w_{i,j}\left(\hat{\mathbf{Y}}_{t,i,j}-\mathbf{Y}_{t,i,j}\right)^2}{\sum_{(i,j)\in\mathcal{S}_t}w_{i,j}}}}{\sigma_w\left(\mathbf{Y}_{t,\mathcal{S}_t}\right)}.
\end{equation}

\noindent Here, $\mathcal{S}_t=\{(i,j):\mathbf{M}_{t,i,j}=0\}$ denotes the spatial indices that are unobserved at time $t$, and $\sigma_w(\cdot)$ is the latitude-weighted spatial standard deviation over these indices.
This normalization ensures the error metric accounts for the varying amplitude of natural variability across seasons and regions.
A lower nRMSE indicates better recovery of spatial anomalies.

\subsection{Reconstruction of four historical climate records}
\label{subsec:observational_inputs}

We used GenCRest to reconstruct historical temperature from HadCRUT5 and Berkeley Earth and precipitation from CRU TS 4.10 and GPCC v2025.
Observational inputs were expressed as monthly anomalies relative to 1961--1990, with unobserved locations left missing before reconstruction.
Within each application, GenCRest-HF and GenCRest-LE shared the same trained model, observational inputs and uncertainty estimates.

Conditional-sampling hyperparameters were calibrated separately for each sampling configuration using the 10\% ERA5 validation subset reserved in a 9:1 training--validation split.
Validation fields were sampled under observation masks from different historical periods and perturbed according to the corresponding observational uncertainties.
These experiments informed a coarse-to-fine parameter search, with the selected configurations summarized in Supplementary Table~A1.

\subsubsection{HadCRUT5 temperature}

We used the non-infilled ensemble-mean monthly anomaly fields from HadCRUT.5.1.0.0, which combine land near-surface air-temperature and ocean sea-surface-temperature (SST) observations on a $5^\circ$ grid~\cite{morice2021updated}.
The primary reconstruction used a unified GenCRest temperature model with sea-ice-masked observational inputs to account for the distinction between SST and air-temperature variability over sea ice.
Using monthly HadISST.2.2.2.0 sea-ice concentrations~\cite{titchner2014met}, we excluded grid-cell months with no land air-temperature contribution and sea-ice concentration exceeding 15\%.
Observations containing a land air-temperature contribution were retained.
The masked regions remained part of the globally complete reconstruction, with their temperature variability inferred from the learned prior conditioned on the remaining observations.

The reported spatially correlated and uncorrelated error components were combined into monthly observational-error covariance matrices.
These matrices were restricted to the retained observations and used in the covariance-weighted observation loss.

\subsubsection{Berkeley Earth temperature}

We combined breakpoint-corrected Berkeley Earth land-station observations with HadSST.4.2.0.0 SST observations~\cite{rohde2020berkeley,rohde_2026_20427092,kennedy2019ensemble}.
Station temperatures were converted to anomalies relative to their 1961--1990 monthly normals.
Missing components of these normals were estimated using an adaptation of the GloSAT procedure~\cite{taylor2025glosat}.

Station anomalies were aggregated independently to sparse $1^\circ$ and $5^\circ$ land grids by averaging observations from unique physical stations within each grid-cell month.
Diagonal land-observation variances combined reported station precision with estimates of reference-period and spatial representativeness uncertainty.

The reconstruction proceeded in two stages.
Before the first stage, $5^\circ$ HadSST observations were removed wherever monthly HadISST2 sea-ice concentration exceeded 15\%, while land observations were retained.
The remaining SST anomalies were blended with the sparse $5^\circ$ land anomalies using grid-cell land--ocean weights, with the available component used where only one was observed.
The blended covariance combined the estimated land variances and reported HadSST spatial error covariance.
These inputs conditioned the trained $5^\circ$ GenCRest temperature model to generate globally complete ensemble members.

In the second stage, the trained $1^\circ$ GenCRest temperature model was conditioned jointly on the sparse $1^\circ$ land observations and the complete $5^\circ$ reconstructions.
Members were paired individually between stages, so that each $1^\circ$ realization used its corresponding $5^\circ$ member.
The land-observation term used the estimated diagonal variances, while a squared-residual term guided the area-averaged $1^\circ$ field toward its paired $5^\circ$ reconstruction.
This stage therefore combined reconstruction from sparse land observations with conditional ocean downscaling from $5^\circ$ to $1^\circ$, requiring ocean structure below the resolution of the marine observations to be inferred from the learned prior.

\subsubsection{CRU TS 4.10 precipitation}

We constructed sparse monthly precipitation fields directly from the CRU TS 4.10 station database~\cite{harris2020version}.
The station-count metadata (\texttt{stn}) accompanying the official gridded product describe the stations contributing to angular-distance-weighted interpolation, which uses a correlation decay distance of 450\,km for precipitation.
These counts therefore do not identify whether observations are available within a particular grid cell.
Working directly with the station records allowed us to retain their spatial distribution and changing monthly availability when constructing the sparse inputs.

Following quality control, precipitation at each station was expressed as a relative anomaly from its corresponding 1961--1990 monthly normal.
Valid station-relative anomalies were averaged within each $0.5^\circ$ grid-cell month and converted to absolute precipitation anomalies using the corresponding CRU TS monthly climatology.
The resulting fields were expressed in $\mathrm{mm\,month^{-1}}$, with grid-cell months lacking valid station observations left missing.
Diagonal observation variances were estimated from within-cell station variance and the number of contributing stations.

\subsubsection{GPCC v2025 precipitation}

We used the $0.5^\circ$ GPCC Precipitation Analysis Monthly Version 2025 and its accompanying gauge counts and interpolation-error estimates~\cite{rustemeier2025gpcc}.
Because the complete underlying station archive is not publicly distributed, we constructed sparse inputs from the published gridded analysis.
GPCC reports the number of gauges located within each grid cell separately from the number contributing to its interpolation.
We used the former to retain only grid-cell months containing at least one reporting gauge, leaving cells without local gauge observations missing.
The retained values thus represent gauge-supported gridded precipitation estimates.

Monthly anomalies were calculated by subtracting the corresponding calendar-month means over 1961--1990, derived from the same GPCC analysis.
The squared published interpolation errors supplied the diagonal observation variances.

Both precipitation applications used the same trained $0.5^\circ$ GenCRest model to reconstruct globally complete precipitation fields.
Conditioning used land observations alone, with precipitation variability over the oceans inferred through the learned spatiotemporal relationships between land and ocean.

\clearpage
\section*{Declarations}
\backmatter

\bmhead{Author contributions}
Z.Q., T.L. and N.Boers conceived and designed the study.
Z.Q. collected and processed the data, developed and implemented GenCRest, designed and performed the experiments, carried out the formal analyses and visualization, and interpreted the results.
T.L. and N. Boers supervised the study and contributed to experimental design and interpretation of the results.
D.C. contributed expertise on observational climate data and climate-field reconstruction, and contributed to the design and interpretation of the validation experiments and manuscript framing.
S.B. contributed climate-science expertise throughout the study, to the interpretation of the reconstruction results and their scientific framing.
N.Bochow advised on the LaMa benchmark.
S.Y., P.H., and C.B. advised on the diffusion modeling and spatial-only benchmark.
M.G. and B.G. contributed to the design and interpretation of the validation analyses and assessment of reconstruction robustness.
Z.Q. wrote the original draft with input from T.L.
All authors contributed to scientific discussions and reviewed and edited the manuscript.

\bmhead{Competing interests} The authors declare no competing interests.

\bmhead{Supplementary information} Supplementary information is available for this paper.

\bmhead{Data availability} 
The temperature and precipitation reconstructions generated with GenCRest will be made publicly available upon acceptance of the manuscript.
The observational and reanalysis datasets used in this study are publicly available from their respective repositories.
HadCRUT.5.1.0.0 data can be accessed at \url{https://www.metoffice.gov.uk/hadobs/hadcrut5/data/HadCRUT.5.1.0.0/download.html}.
CRUTEM.5.1.0.0 data can be accessed at \url{https://www.metoffice.gov.uk/hadobs/crutem5/data/CRUTEM.5.1.0.0/download.html}.
HadSST.4.2.0.0 data can be accessed at \url{https://www.metoffice.gov.uk/hadobs/hadsst4/data/download.html}.
HadISST.2.2.2.0 data can be accessed at \url{https://www.metoffice.gov.uk/hadobs/hadisst2/data/download.html}.
Berkeley Earth Temperature data are available at \url{https://berkeleyearth.org/data/}.
Berkeley Earth merged weather stations are available at \url{https://zenodo.org/records/20427092}.
CRU TS 4.10 data are hosted at \url{https://crudata.uea.ac.uk/cru/data/hrg/cru_ts_4.10/}.
GPCC v2025 precipitation data can be downloaded from \url{https://opendata.dwd.de/climate_environment/GPCC/html/gpcc_precipitation_analysis_monthly_v2025_doi_download.html}.
GPCP v2.3 precipitation data are available via the NOAA Physical Sciences Laboratory at \url{https://psl.noaa.gov/data/gridded/data.gpcp.html}.
ERA5 reanalysis data can be downloaded from the Copernicus Climate Change Service (C3S) Climate Data Store at \url{https://cds.climate.copernicus.eu/datasets/reanalysis-era5-single-levels?tab=download}.
JRA3Q reanalysis data can be downloaded from the Japan Meteorological Agency at \url{https://www.data.jma.go.jp/jra/html/JRA-3Q/index_en.html}.
NOAA 20CR v3 reanalysis data can be downloaded from the NOAA Physical Sciences Laboratory at \url{https://psl.noaa.gov/data/20thC_Rean/}.
CMIP6 historical simulations from ACCESS-CM2, BCC-CSM2-MR, CESM2, EC-Earth3, FGOALS-f3-L, GFDL-ESM4, MIROC6, MPI-ESM1-2-HR, MRI-ESM2-0, NorESM2-MM and CMCC-ESM2, together with HighResMIP simulations from CNRM-CM6-1-HR, ECMWF-IFS-HR, GFDL-CM4C192 and HadGEM3-GC31-HM, were accessed as cloud-optimized Zarr datasets through the Pangeo CMIP6 catalog hosted on Google Cloud Storage (\url{https://storage.googleapis.com/cmip6/pangeo-cmip6.json}).
The underlying model outputs were originally published through the Earth System Grid Federation.
NOAA Climate Data Record (CDR) of AVHRR Polar Pathfinder Extended (APP-X) can be downloaded from the NOAA National Centers for Environmental Information at \url{https://www.ncei.noaa.gov/data/avhrr-polar-pathfinder-extended/access/nhem/}.

\bmhead{Code availability} The code for GenCRest and the associated analyses will be publicly available upon acceptance of this manuscript for publication.

\bmhead{Acknowledgements} T.L. acknowledges funding from the National Key R\&D Program of China no.2023YFE0109000.
Z.Q. acknowledges funding from the program of the China Scholarship Council.
N.Boers and S.B. acknowledge funding from the Volkswagen Foundation.
N.Bochow received funding from the European Union (ERC, FORCLIMA, 101044247).
This is ClimTip contribution \#138; the ClimTip project has received funding from the European Union’s Horizon Europe research and innovation programme under grant agreement no. 101137601.
This study received support from the European Space Agency Climate Change Initiative (ESA-CCI) Tipping Elements SIRENE project (contract no. 4000146954/24/I-LR).
The authors gratefully acknowledge the Ministry of Research, Science and Culture (MWFK) of Land Brandenburg for supporting this project by providing resources on the high-performance computer system at the Potsdam Institute for Climate Impact Research.
\bibliography{ref}

\clearpage

\setcounter{figure}{0}
\renewcommand{\theHfigure}{extended.\arabic{figure}}
\renewcommand{\figurename}{Extended Data Fig.}

\begin{figure}[!htbp]
\centering
\includegraphics[width=\linewidth]{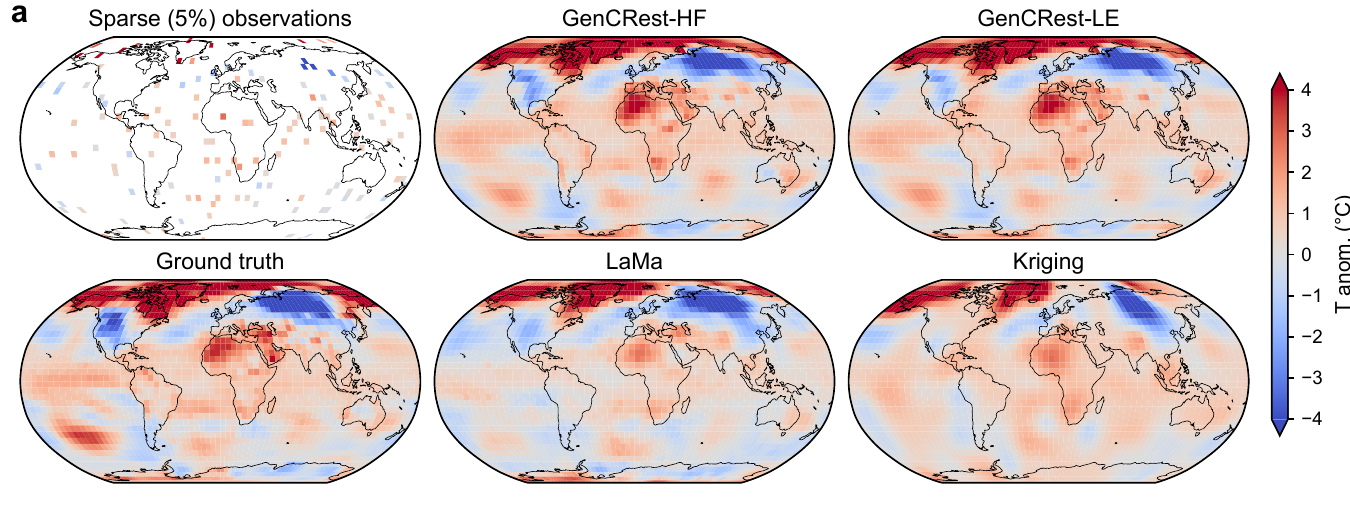}
\includegraphics[width=\linewidth]{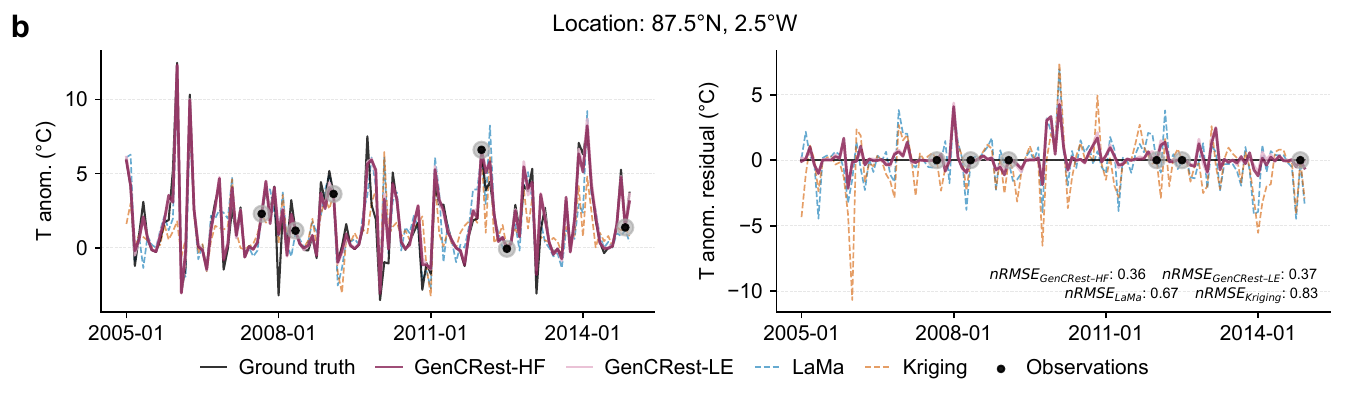}
\includegraphics[width=\linewidth]{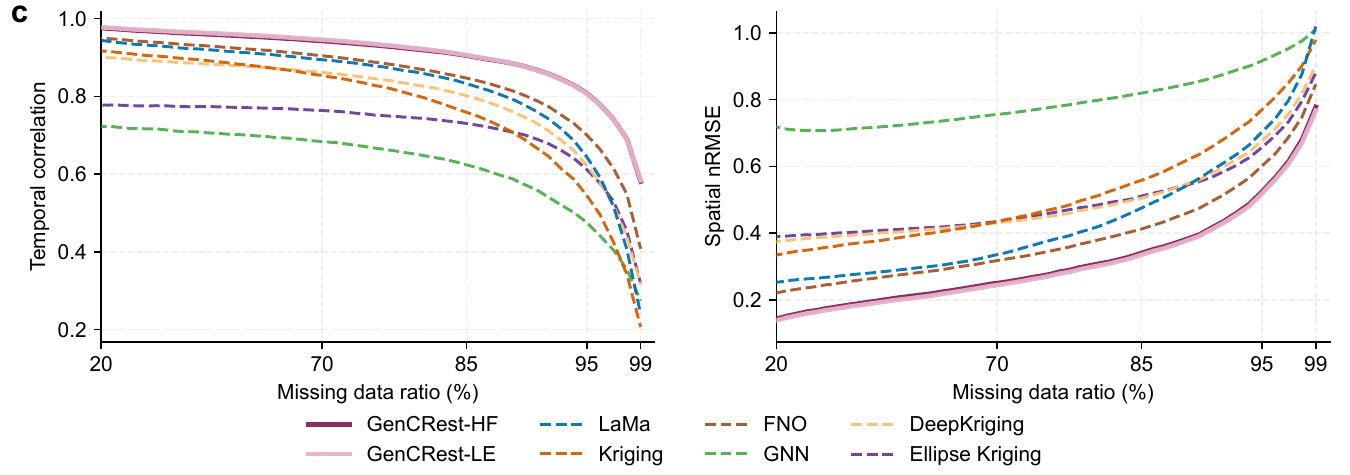}
\caption{
\textbf{Idealized, noise-free gap filling of global temperature.}
Complete ERA5 fields are randomly subsampled, with retained values treated as exact observations and withheld values used as the evaluation reference.
\textbf{a}, Representative monthly $5^\circ$ temperature-anomaly fields reconstructed from 5\% observational coverage using GenCRest-HF, GenCRest-LE, LaMa, and Kriging.
\textbf{b}, Reconstructed time series and residuals, defined as reconstruction minus ERA5, for the $5^\circ$ grid cell centered at $87.5^\circ$N, $2.5^\circ$W.
Black dots denote observations; annotations give temporal nRMSE.
\textbf{c}, Reconstruction skill across the tested missing-data ratios, evaluated only at unobserved grid cells in each month using the area-weighted mean temporal correlation and time-averaged spatial nRMSE.
Additional benchmarks are a Fourier neural operator (FNO), graph neural network (GNN), DeepKriging and ellipse Kriging.
All maps, time series and skill metrics use ensemble-mean reconstructions from $N=10$ members for GenCRest-HF and $N=50$ members for GenCRest-LE.
Supplementary Section~B.1 reports additional temperature diagnostics, with the benchmark fields and time series shown in Supplementary Fig.~B4.
}
\label{fig:temp5_random}
\end{figure}

\clearpage
\begin{figure}[!htbp]
\centering
\includegraphics[width=\linewidth]{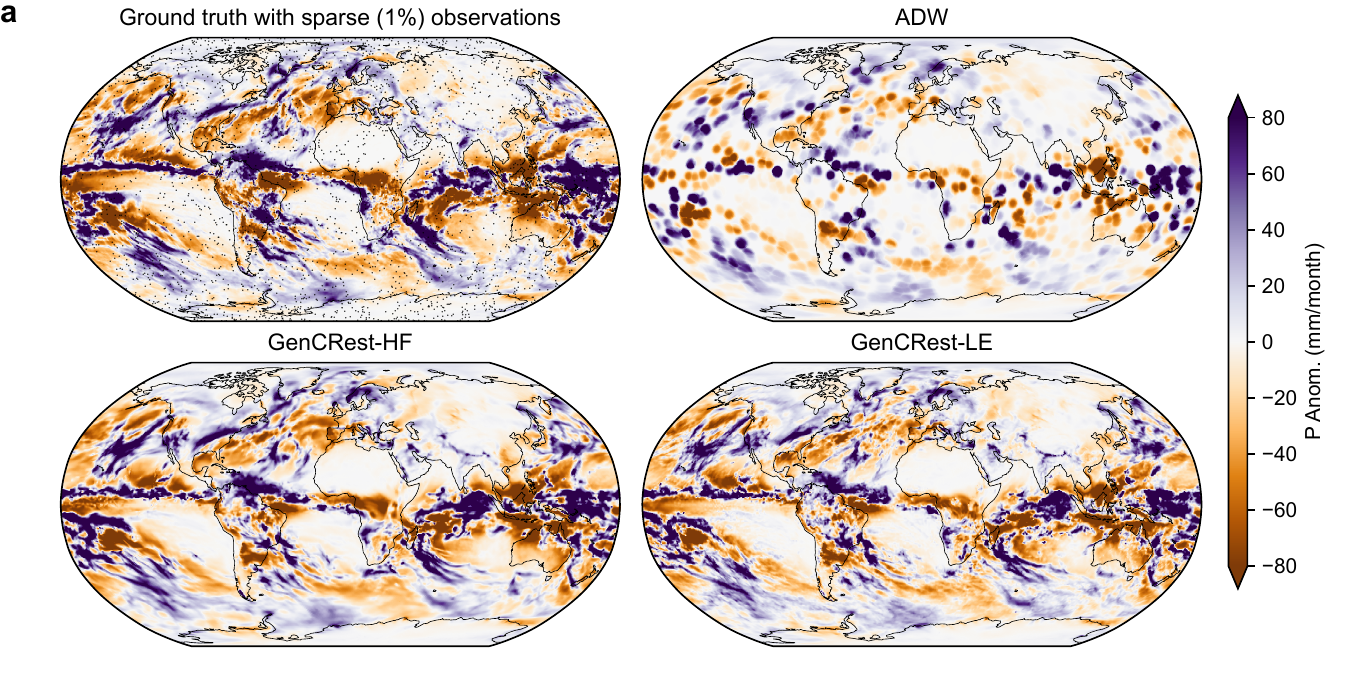}
\includegraphics[width=\linewidth]{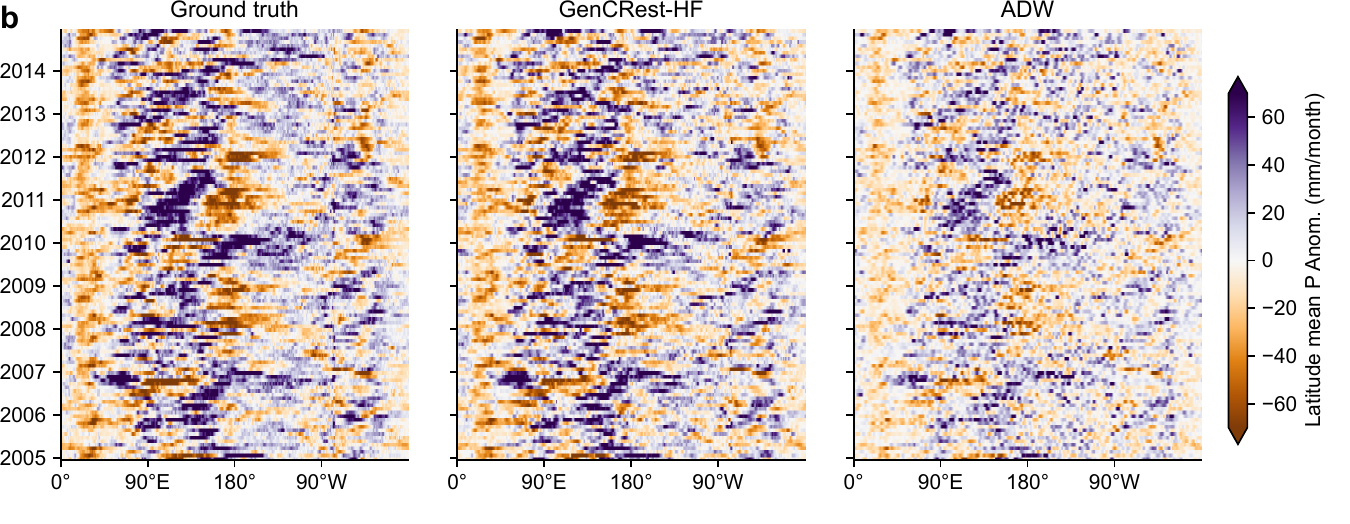}
\includegraphics[width=\linewidth]{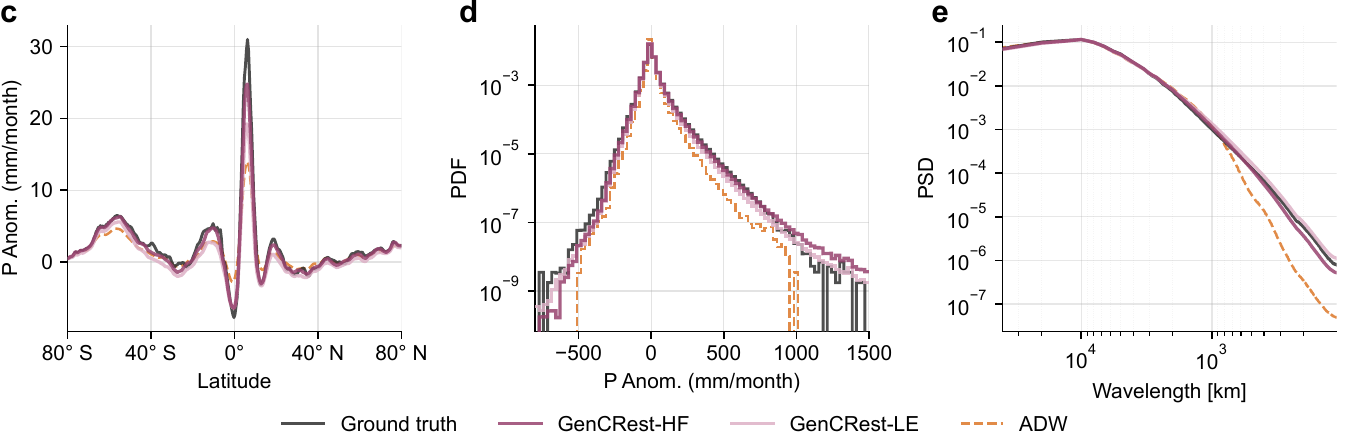}
\caption{
\textbf{Idealized, noise-free gap filling of global precipitation at $0.5^\circ$ resolution.}
ERA5 test fields are randomly subsampled at 1\% coverage, with retained values treated as exact observations.
\textbf{a}, Representative monthly ground truth with sparse observations overlaid as black dots, the ADW reconstruction, and the first members of the GenCRest-HF and GenCRest-LE ensembles.
\textbf{b}, Hovm\"oller diagrams of precipitation anomalies averaged over $15^\circ$S--$15^\circ$N, comparing ERA5 with the first GenCRest-HF member and ADW.
\textbf{c}, Zonal-mean precipitation anomalies averaged over the test period.
\textbf{d}, Spatial probability density functions (PDFs) calculated for each month and averaged over time.
\textbf{e}, Temporally averaged spatial power spectral density (PSD), showing the distribution of power across spatial scales.
For \textbf{c--e}, each diagnostic is calculated separately for each member and then averaged across 10 GenCRest-HF members and 50 GenCRest-LE members.
This preserves member-level distributional and spectral information when summarizing the ensembles.
ADW is deterministic.
The LaMa comparison and reconstruction from land observations are shown in Supplementary Figs.~B6 and~B7, respectively.
}
\label{fig:precip05_random}
\end{figure}

\clearpage
\begin{figure}[!htbp]
\centering
\includegraphics[width=.98\linewidth]{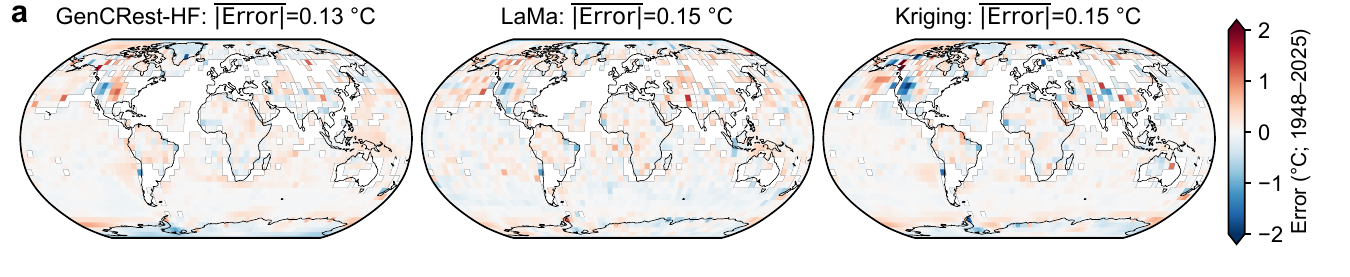}
\includegraphics[width=.98\linewidth]{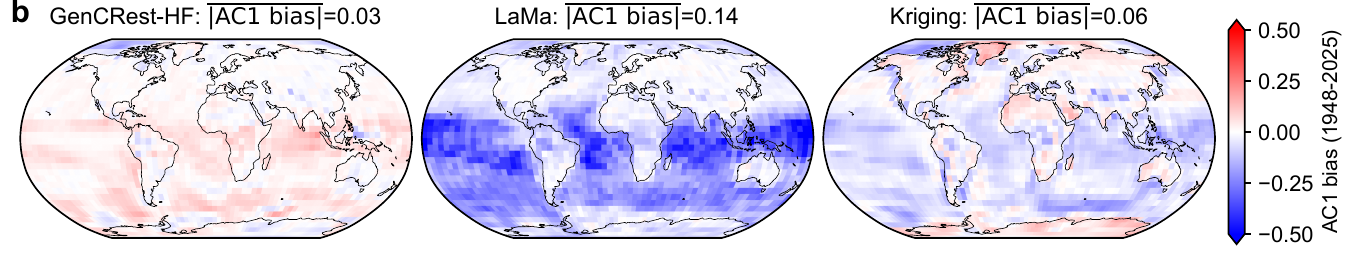}
\includegraphics[width=.98\linewidth]{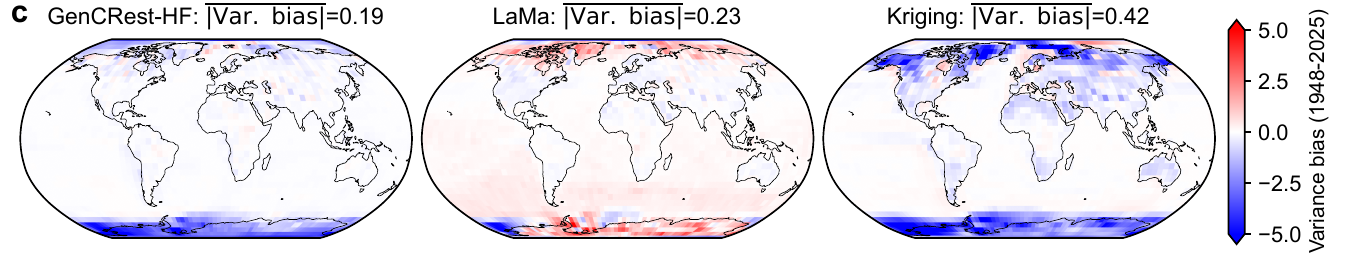}
\includegraphics[width=.98\linewidth]{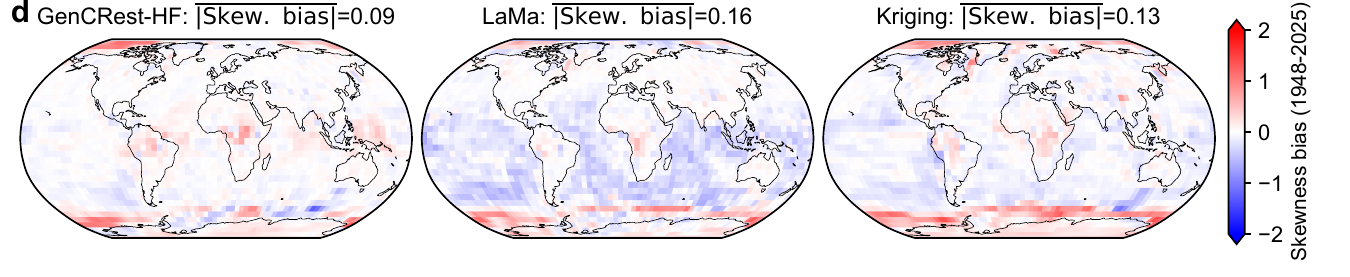}
\includegraphics[width=.98\linewidth]{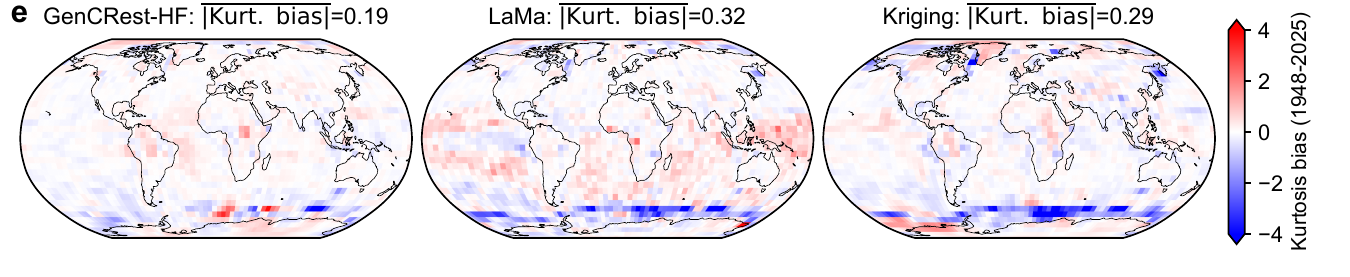}
\includegraphics[width=.98\linewidth]{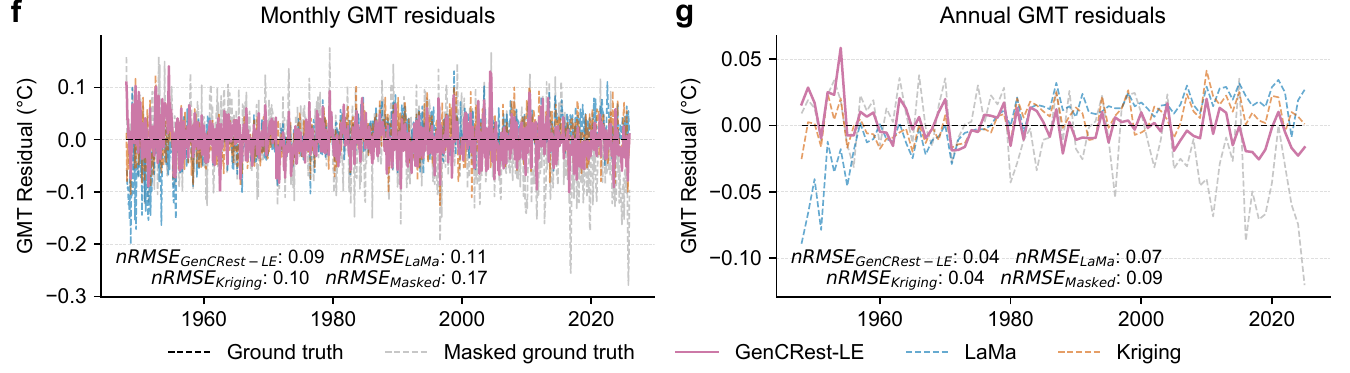}
\caption{
\textbf{Temperature reconstruction under contemporaneous (1948--2025) HadCRUT5 coverage and observational uncertainty.}
$5^\circ$ JRA-3Q fields over 1948--2025 are sampled using the temporally corresponding monthly HadCRUT5 masks and perturbed using the corresponding observational-error covariance matrices.
The complete, unperturbed JRA-3Q sequence serves as the evaluation reference.
Columns show GenCRest-HF, LaMa and Kriging, respectively.
\textbf{a}, Time-mean signed reconstruction errors calculated at each grid cell over unobserved months; grid cells observed throughout the evaluation period are left blank.
\textbf{b--e}, Biases in AC1, temporal variance, skewness and kurtosis, respectively, calculated from complete reconstructed and reference time series.
AC1 is calculated after removing a linear trend; variance, skewness and excess kurtosis are calculated from the undetrended monthly anomaly series.
Errors and biases are defined as reconstructed values or statistics minus their JRA-3Q counterparts, with red and blue indicating positive and negative values.
GenCRest-HF fields are averaged across 10 members before calculating these diagnostics.
Map annotations denote area-weighted mean absolute grid-cell errors or statistical biases and correspond to the 1948--2025 entries in Fig.~3d of the main text.
\textbf{f}, \textbf{g}, Monthly and annual GMT residuals, respectively, defined as reconstructed GMT minus the GMT of the complete, unperturbed JRA-3Q reference.
Results are shown for the 50-member GenCRest-LE ensemble mean and the benchmark reconstructions.
The masked-reference curve shows the corresponding GMT residuals calculated from the incomplete, perturbed pseudo-observations relative to the same JRA-3Q reference.
}
\label{fig:temp5_jra3q_highstat}
\end{figure}

\clearpage
\begin{figure}[!htbp]
\centering
\includegraphics[width=1.\linewidth]{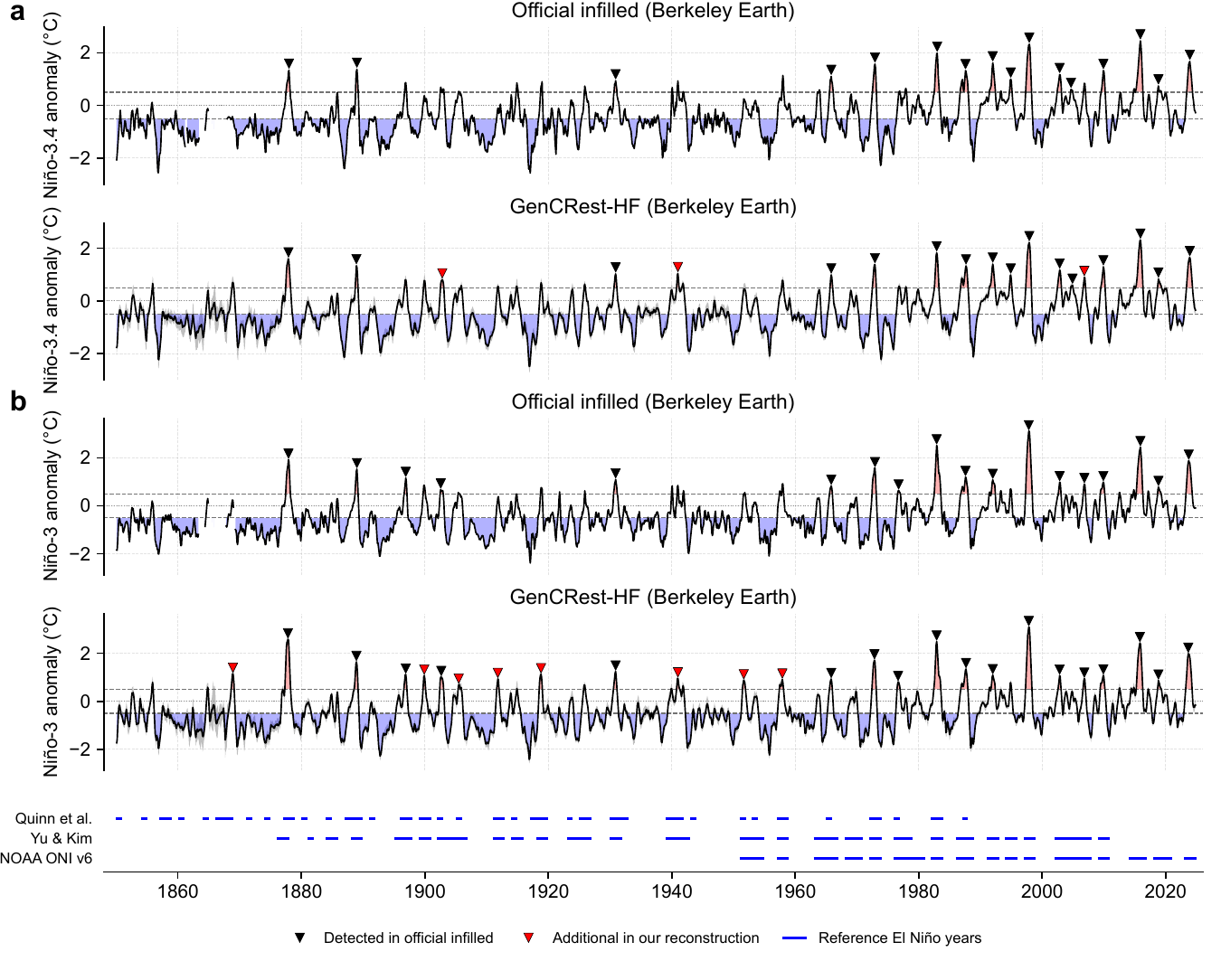}
\includegraphics[width=1.\linewidth]{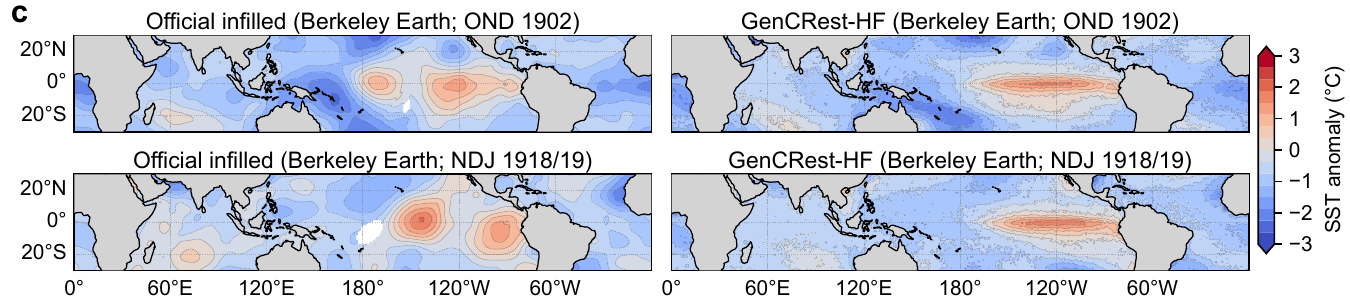}
\caption{
\textbf{Historical El Ni\~no variability in Berkeley Earth temperature reconstructions.}
\textbf{a}, Ni\~no~3.4 indices ($5^\circ\mathrm{S}$--$5^\circ\mathrm{N}$, $170^\circ\mathrm{W}$--$120^\circ\mathrm{W}$) for the official infilled product (top) and GenCRest-HF reconstruction (bottom) over 1850--2024.
\textbf{b}, Corresponding Ni\~no~3 indices ($5^\circ\mathrm{S}$--$5^\circ\mathrm{N}$, $150^\circ\mathrm{W}$--$90^\circ\mathrm{W}$).
Both indices use cosine-latitude weighting and a centered three-month running mean.
Reconstructed indices show ensemble means, with gray shading spanning the member minimum and maximum.
Red and blue shading highlights anomalies above $0.5\,{}^\circ\mathrm{C}$ and below $-0.5\,{}^\circ\mathrm{C}$, respectively.
Warm episodes require the relevant index to reach at least $0.5\,{}^\circ\mathrm{C}$ for at least five consecutive overlapping three-month seasons.
Ni\~no~3.4 provides the primary El Ni\~no classification, while Ni\~no~3 characterizes the eastern-Pacific expression of warm episodes.
Triangles mark peak seasons.
All episodes detected in the official indices are marked in black; in the reconstruction panels, black and red triangles distinguish episodes with and without temporal overlap with episodes in the corresponding official index.
The shared reference timeline shows event-year ranges from the independent historical compilation of Quinn et al.~\cite{quinn1987nino} (confidence ratings $\geq 4$) and the SST-based records of Yu and Kim~\cite{yu2012identifying} and NOAA ONI v6~\cite{NOAA_ONI}.
Blue bars indicate calendar-year ranges and are shown irrespective of detection in the plotted indices.
The reconstructed 1918--1919 episode meets the persistence criterion in Ni\~no~3 but not in Ni\~no~3.4, where only three consecutive seasons reach the threshold.
\textbf{c}, Three-month mean sea-surface temperature anomalies during October--December 1902 (top) and November 1918--January 1919 (bottom).
Columns show the official infilled product (left) and the GenCRest-HF ensemble mean (right).
All anomalies are relative to 1991--2020.
The corresponding HadCRUT5 analysis is shown in Fig.~\ref{fig:temp_recon}b,c.
}
\label{fig:berkeley_enso}
\end{figure}

\clearpage
\begin{figure}[!htbp]
\centering
\includegraphics[width=\linewidth]{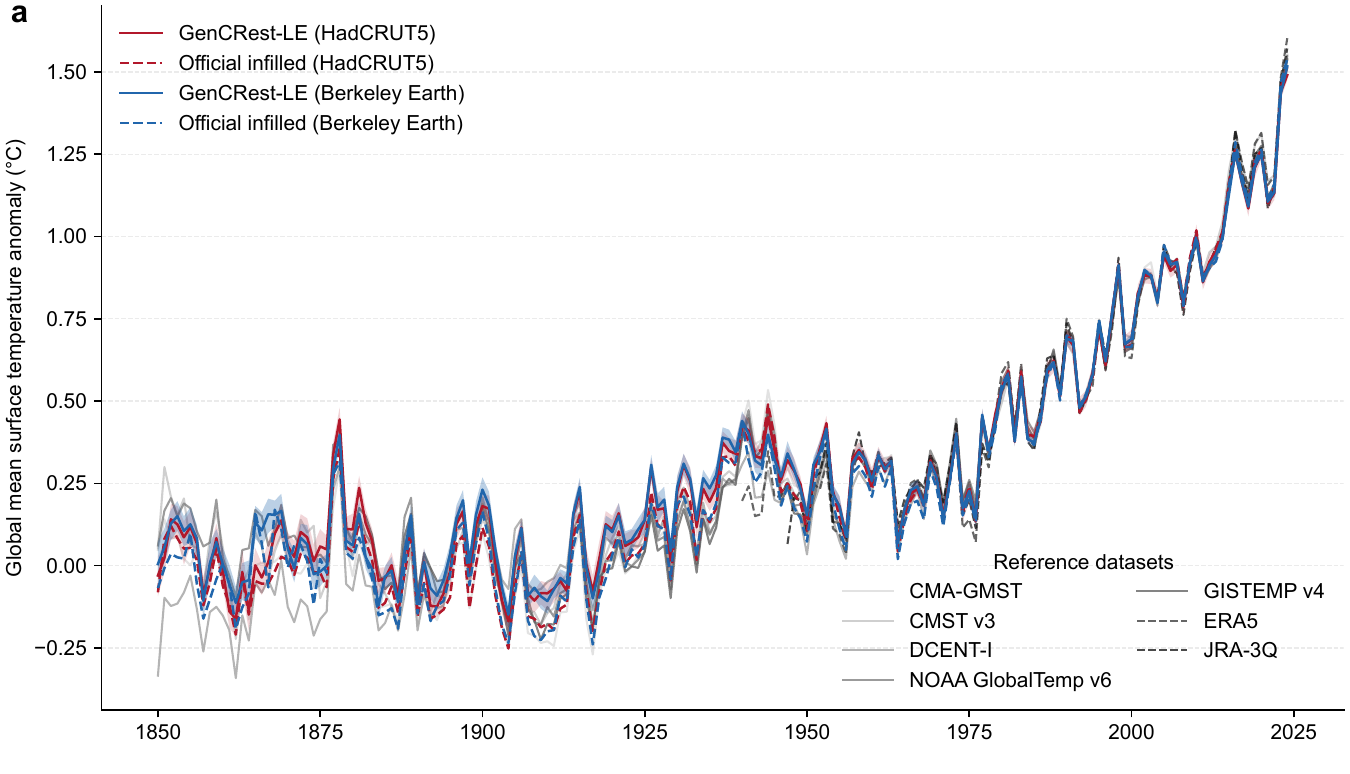}
\includegraphics[width=\linewidth]{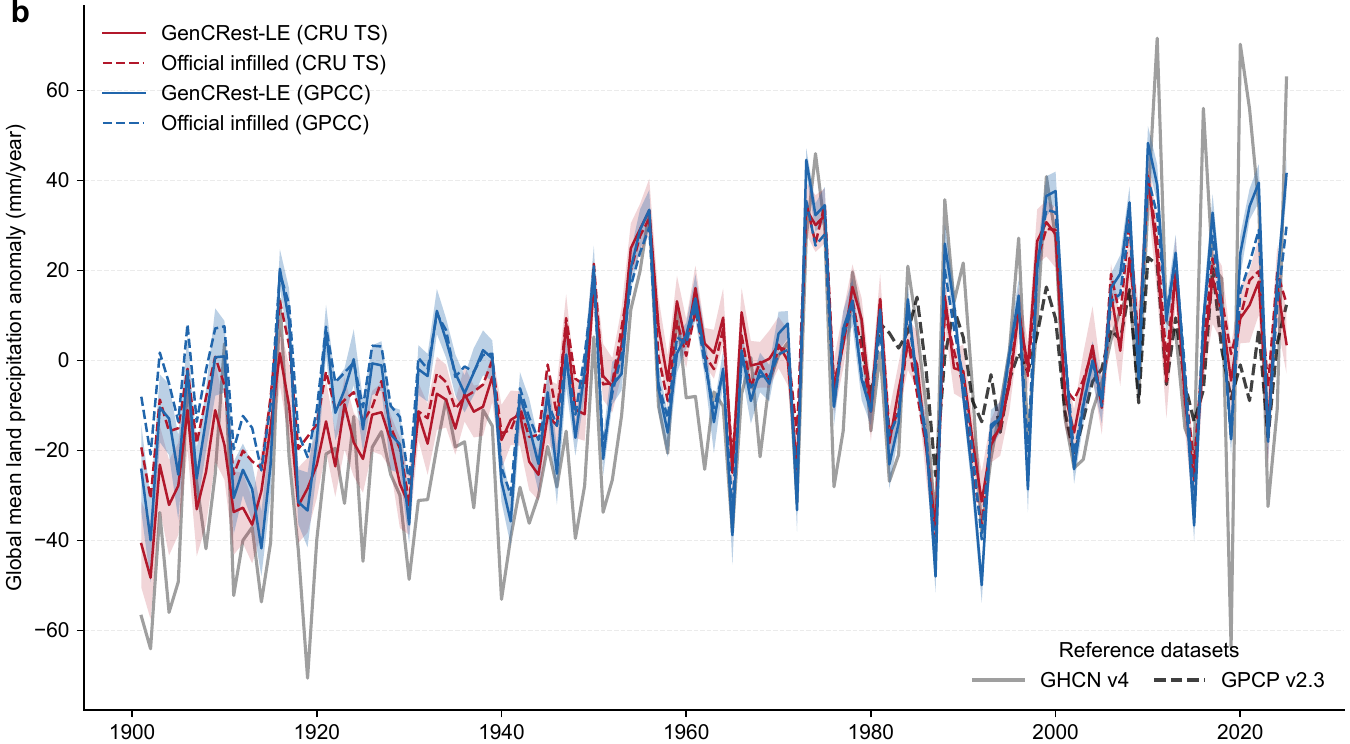}
\caption{
\textbf{Global mean temperature and land precipitation across reconstructions and reference products.}
\textbf{a}, Annual global mean surface-temperature anomalies from the HadCRUT5 and Berkeley Earth reconstructions over 1850--2024, compared with their official infilled products, CMA-GMST, CMST v3, DCENT-I, NOAA GlobalTemp v6, GISTEMP v4, ERA5, and JRA-3Q.
Annual temperature series are rebased to 1981--2010 and shifted by $+0.69\,{}^\circ\mathrm{C}$.
\textbf{b}, Annual global land-mean precipitation anomalies from the CRU TS 4.10 and GPCC v2025 reconstructions over 1901--2025, compared with their official infilled products, GHCN v4, and GPCP v2.3.
All precipitation series use the 1981--2010 anomaly reference period.
For reconstructed fields, monthly spatial means use cosine-latitude weights over valid grid cells; annual temperature values average these monthly means, whereas annual precipitation values sum the monthly totals.
Each reconstruction uses 50 GenCRest-LE members.
Solid red and blue curves show the means of the annual member estimates, and shading indicates their 2.5th--97.5th percentile range.
Corresponding dashed curves show the official infilled products.
Additional reference products appear in gray or black, with line styles identified in the legends.
}
\label{fig:global_mean}
\end{figure}

\clearpage
\begin{figure}[!htbp]
\centering
\includegraphics[width=\linewidth]{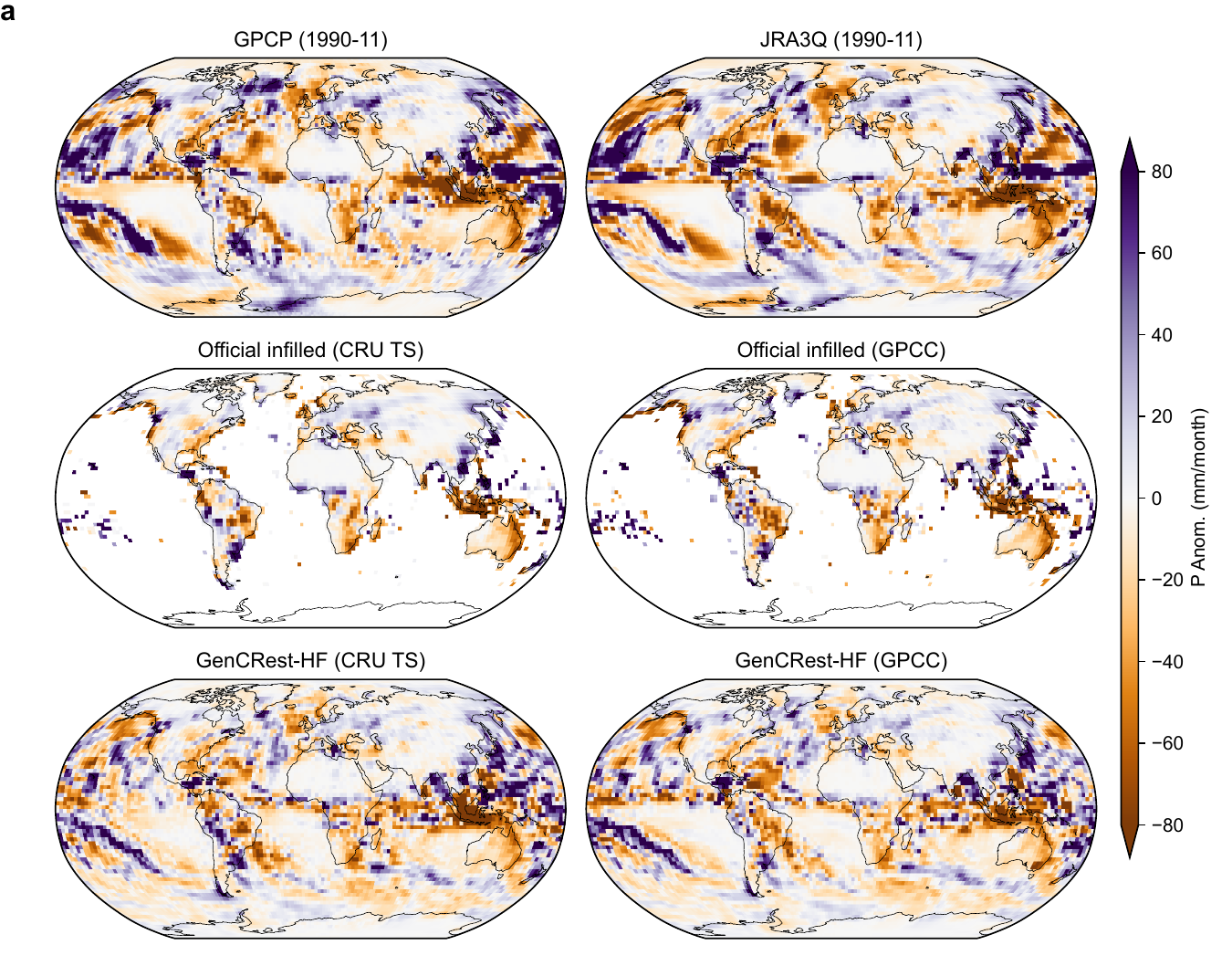}
\includegraphics[width=\linewidth]{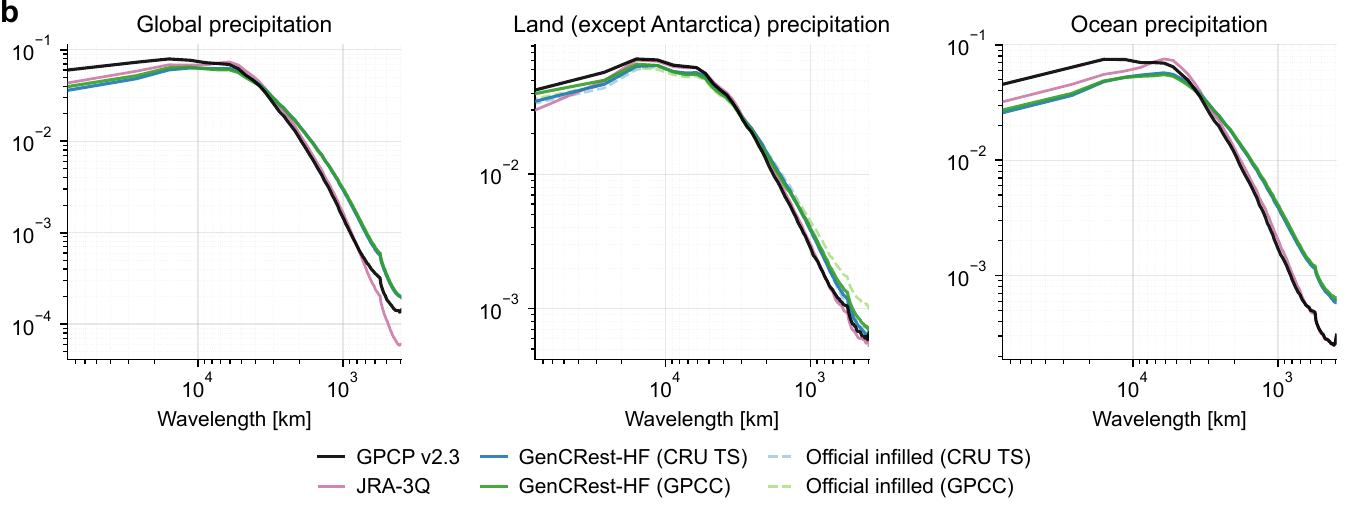}
\caption{
\textbf{Reconstructed precipitation structure and normalized spatial spectra compared with satellite and reanalysis products.}
\textbf{a}, November 1990 precipitation anomalies from GPCP v2.3 and JRA-3Q (top), the official CRU TS 4.10 and GPCC v2025 products (middle), and single GenCRest-HF reconstructions conditioned on the corresponding land observations (bottom).
All fields are compared on a common $2.5^\circ$ grid, with the $0.5^\circ$ fields conservatively regridded to this resolution.
Anomalies are relative to 1981--2010 and expressed in $\mathrm{mm\,month^{-1}}$.
\textbf{b}, Normalized spatial power spectral densities (PSDs) for global, land, and ocean precipitation, averaged over 1980--2025.
Each monthly spectrum is normalized to unit summed power before temporal averaging.
Reconstruction spectra are calculated separately for 10 GenCRest-HF members and then averaged.
Solid black, pink, blue, and green curves denote GPCP v2.3, JRA-3Q, the CRU TS reconstruction, and the GPCC reconstruction, respectively.
Pale dashed curves in the land comparison show the corresponding official infilled products.
}
\label{fig:precip_gpcp_jra3q}
\end{figure}

\clearpage
 \begin{figure}[htbp]
\centering
\includegraphics[width=.98\linewidth]{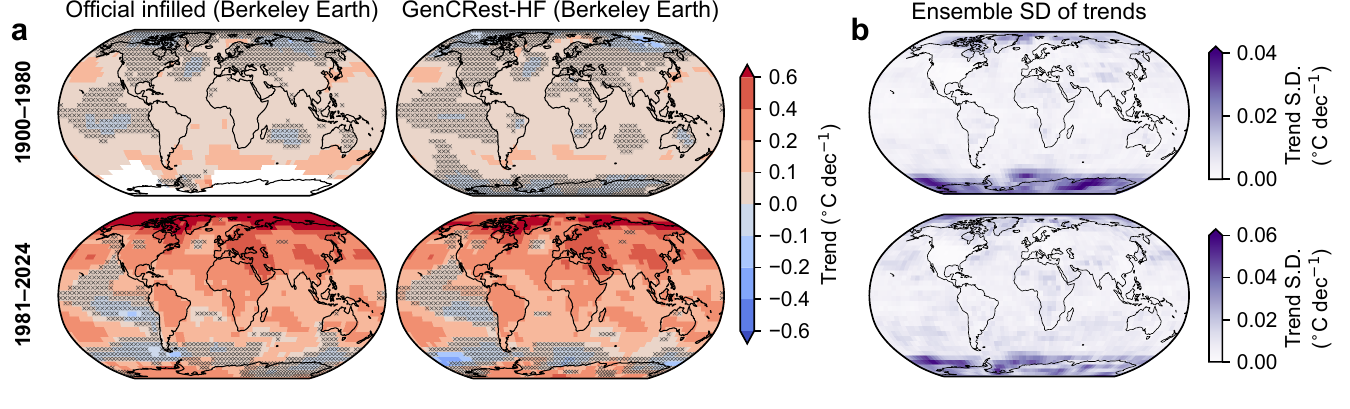}
\includegraphics[width=.49\linewidth]{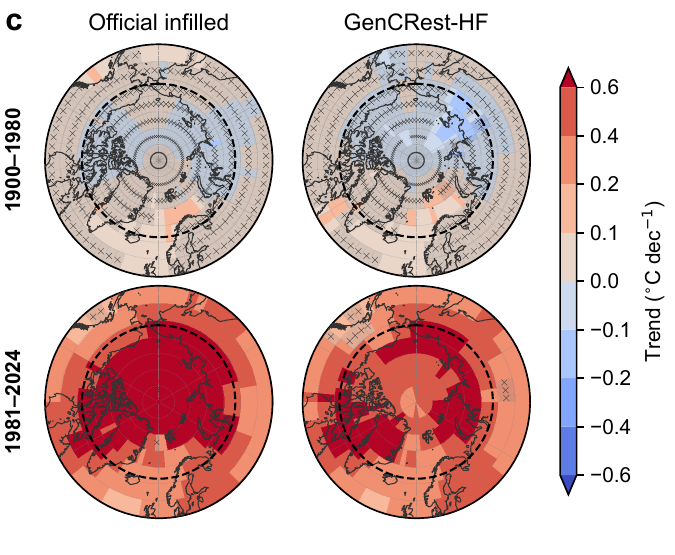}
\includegraphics[width=.49\linewidth]{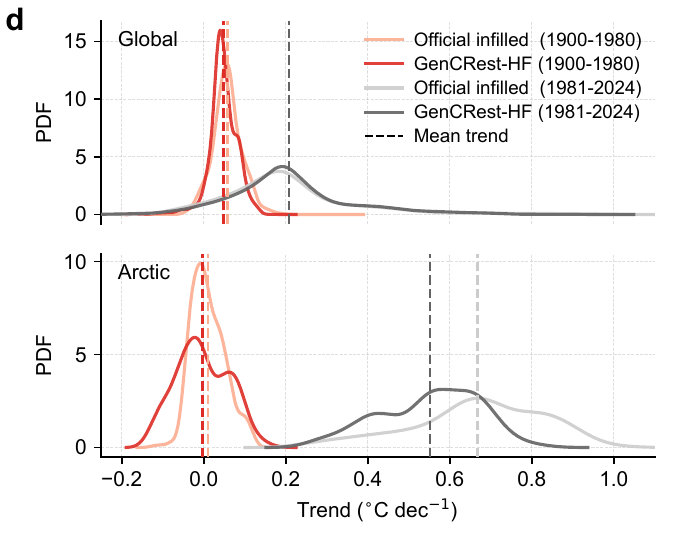}
\includegraphics[width=.98\linewidth]{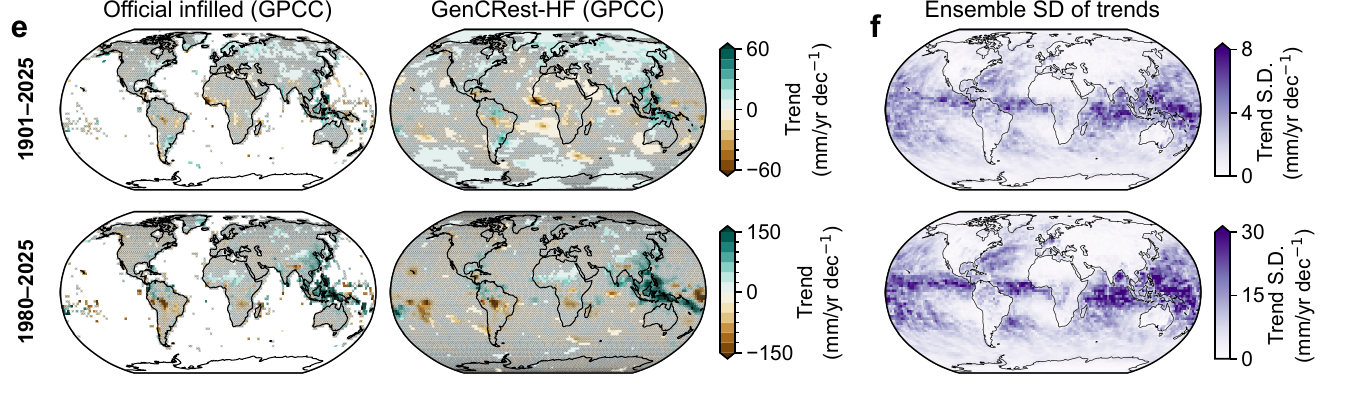}
\includegraphics[width=.98\linewidth]{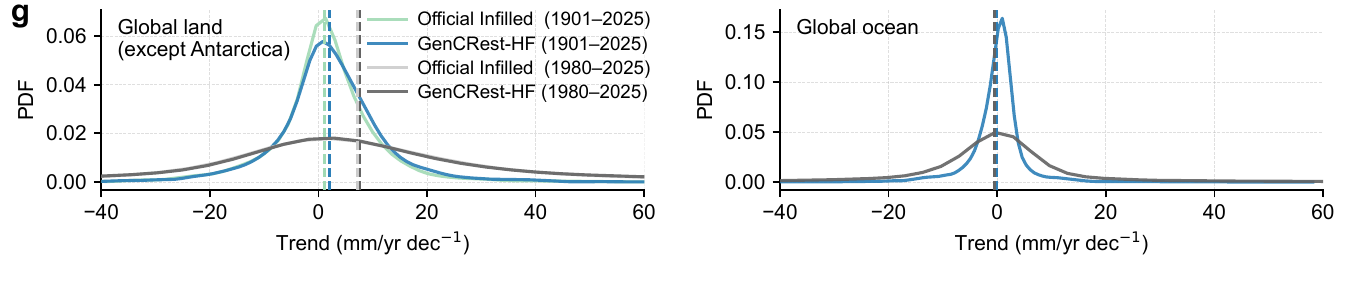}
\caption{
\textbf{Historical temperature and precipitation trends reconstructed from Berkeley Earth and GPCC.}
Temperature and precipitation fields are aggregated to $5^\circ$ and $2.5^\circ$, respectively, before calculating trends and ensemble spread.
\textbf{a}, Linear temperature trends in the Berkeley Earth official infilled product and GenCRest-HF reconstruction during 1900--1980 (top) and 1981--2024 (bottom).
Reconstructed temperature trends are calculated from the 10-member ensemble-mean field.
\textbf{b}, Ensemble spread, expressed as the standard deviation of trends calculated separately for each reconstruction member.
\textbf{c}, Corresponding Arctic temperature trends, with dashed circles marking $66.5^\circ$\,N.
\textbf{d}, Area-weighted probability density functions (PDFs) of temperature trends across the globe (top) and the Arctic (bottom).
The shift toward weaker Arctic warming during 1981--2024 also reflects differences in observational inputs, with generally lower warming trends already present in Berkeley Earth's station records and our aggregated fields than in the official infilled product (Supplementary Fig.~B40).
\textbf{e}, Linear trends in annual precipitation from the GPCC v2025 official infilled product and GenCRest-HF reconstruction during 1901--2025 (top) and 1980--2025 (bottom).
Reconstructed precipitation trends are averaged across 10 members.
\textbf{f}, Corresponding ensemble spread in precipitation trends, calculated as in \textbf{b}.
\textbf{g}, Area-weighted PDFs of precipitation trends over global land excluding Antarctica (left) and of reconstructed trends over the global ocean (right).
The distributions in \textbf{d} and \textbf{g} include all valid trends, with vertical dashed lines indicating their area-weighted means.
Each precipitation distribution in \textbf{g} uses the valid grid cells available for that field within the stated land or ocean domain.
Gray crosses in \textbf{a}, \textbf{c} and \textbf{e} indicate trends that do not meet the significance criterion used in Fig.~\ref{fig:largescale_trend}.
}
\label{fig:berkeley_gpcc_largescale_trend}
\end{figure}

\clearpage
 \begin{figure}[htbp]
\centering
\includegraphics[width=.45\linewidth]{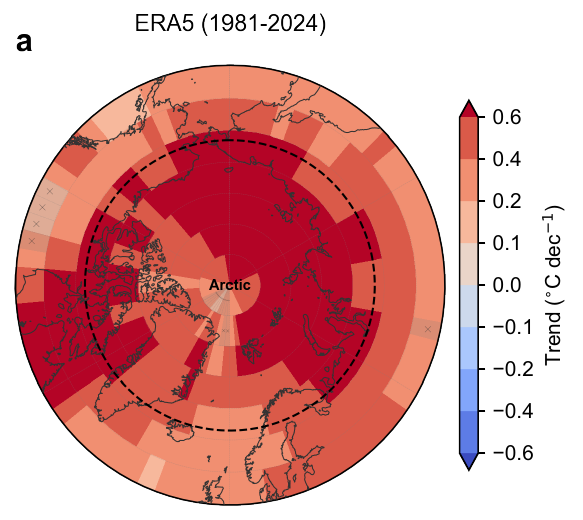}
\includegraphics[width=.45\linewidth]{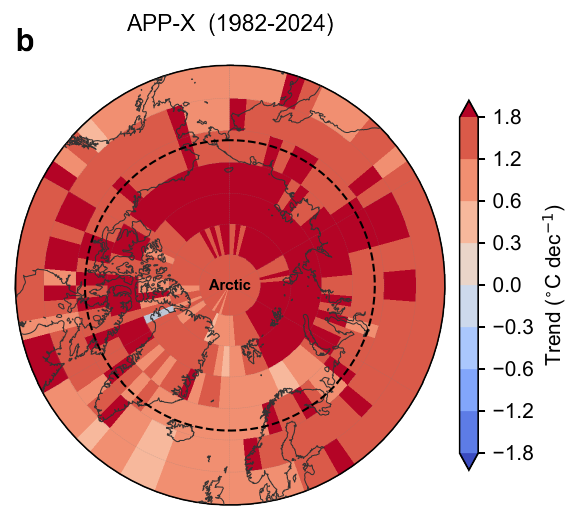}
\includegraphics[width=\linewidth]{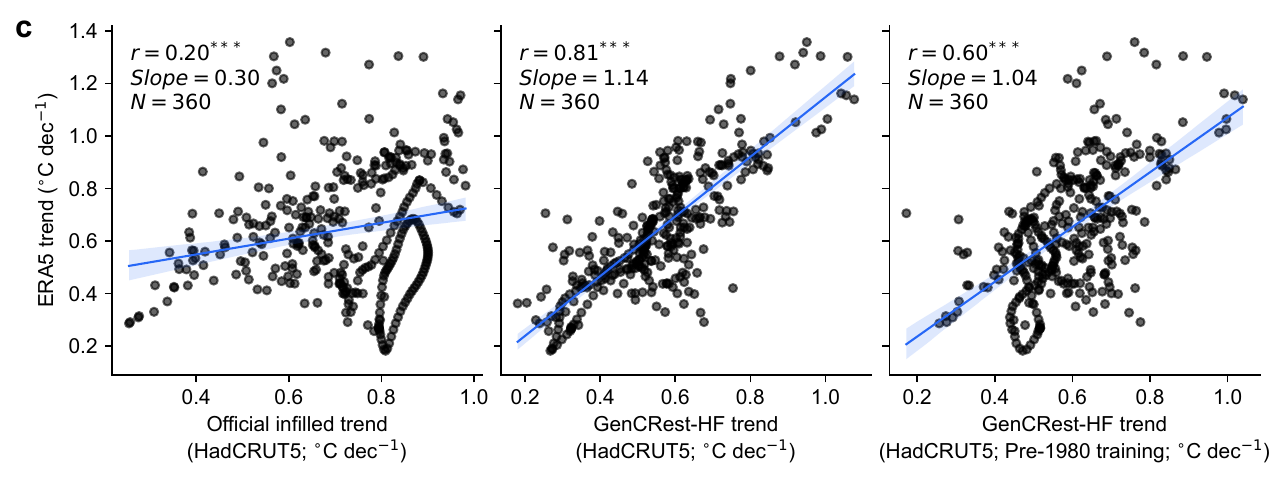}
\includegraphics[width=\linewidth]{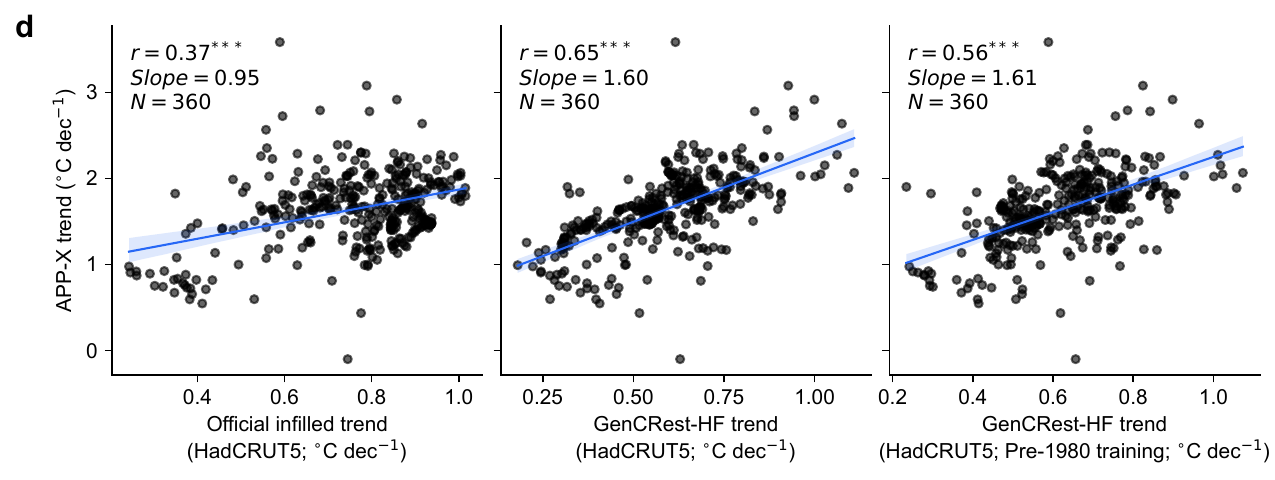}
\caption{
\textbf{Arctic warming patterns compared with reanalysis and satellite observations.}
\textbf{a}, ERA5 near-surface air-temperature trends over 1981--2024.
\textbf{b}, APP-X satellite surface skin-temperature trends over 1982--2024.
All fields are evaluated on a common $5^\circ$ grid.
Dashed circles mark $66.5^\circ$\,N, and gray crosses indicate trends that do not meet the significance criterion used in Fig.~\ref{fig:largescale_trend}.
\textbf{c}, \textbf{d}, Grid-cell specific trend comparison with ERA5 and APP-X, respectively, over their corresponding reference periods.
Columns show the HadCRUT5 official infilled product (left), the full-training GenCRest-HF reconstruction (middle), and the reconstruction with post-1980 climate-model and reanalysis fields excluded from training (right).
Both reconstructions use identical observational conditions, with trends calculated from their respective 10-member ensemble means.
Each point represents an Arctic grid cell.
Blue lines show ordinary least-squares fits, with shading indicating pointwise 95\% confidence intervals obtained from 1,000 bootstrap resamples.
Annotations report Pearson spatial correlations ($r$), regression slopes and sample sizes ($N$); $^{***}$ denotes $p<0.001$.
Corresponding Berkeley Earth and DCENT comparisons are shown in Supplementary Figs.~B43 and~B44.
}
\label{fig:temp5_arctic_reference}
\end{figure}

\clearpage
 \begin{figure}[htbp]
\centering
\includegraphics[width=\linewidth]{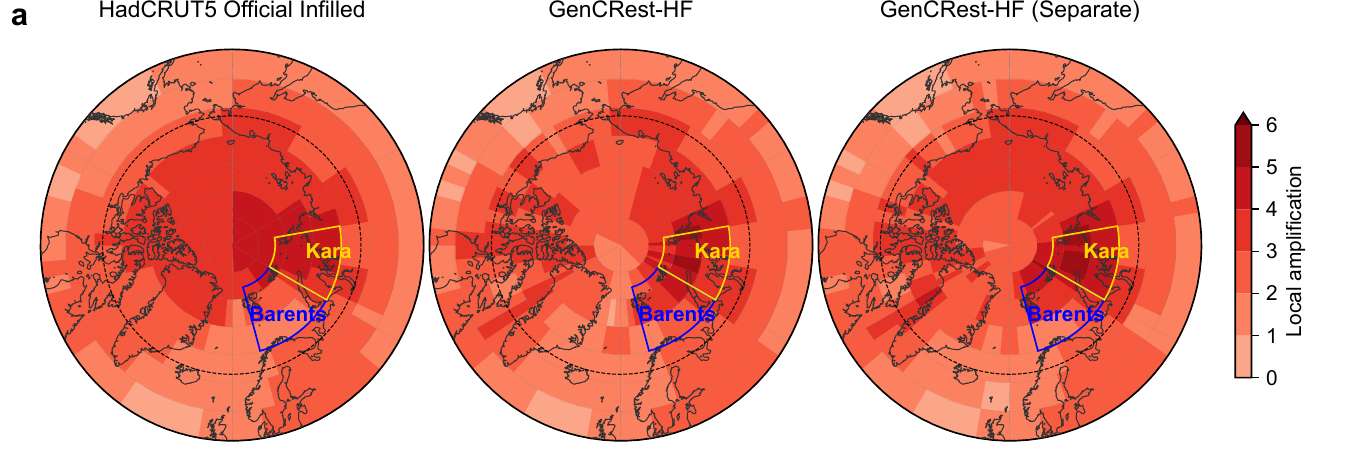}
\includegraphics[width=\linewidth]{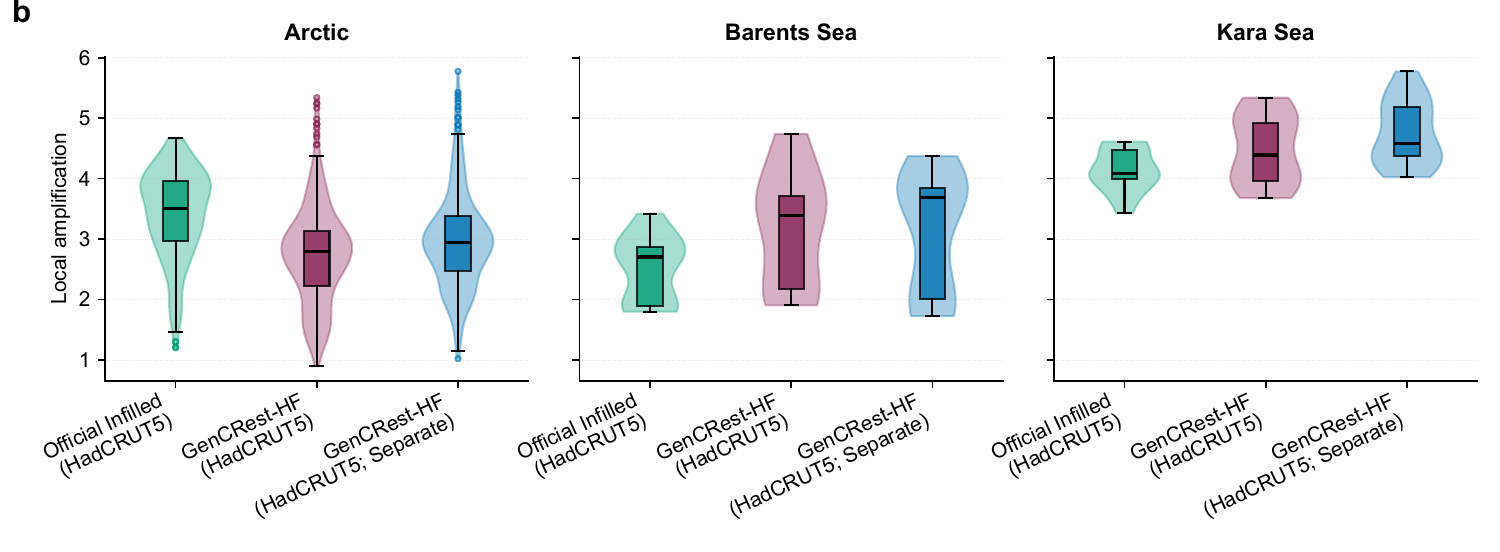}
\includegraphics[width=\linewidth]{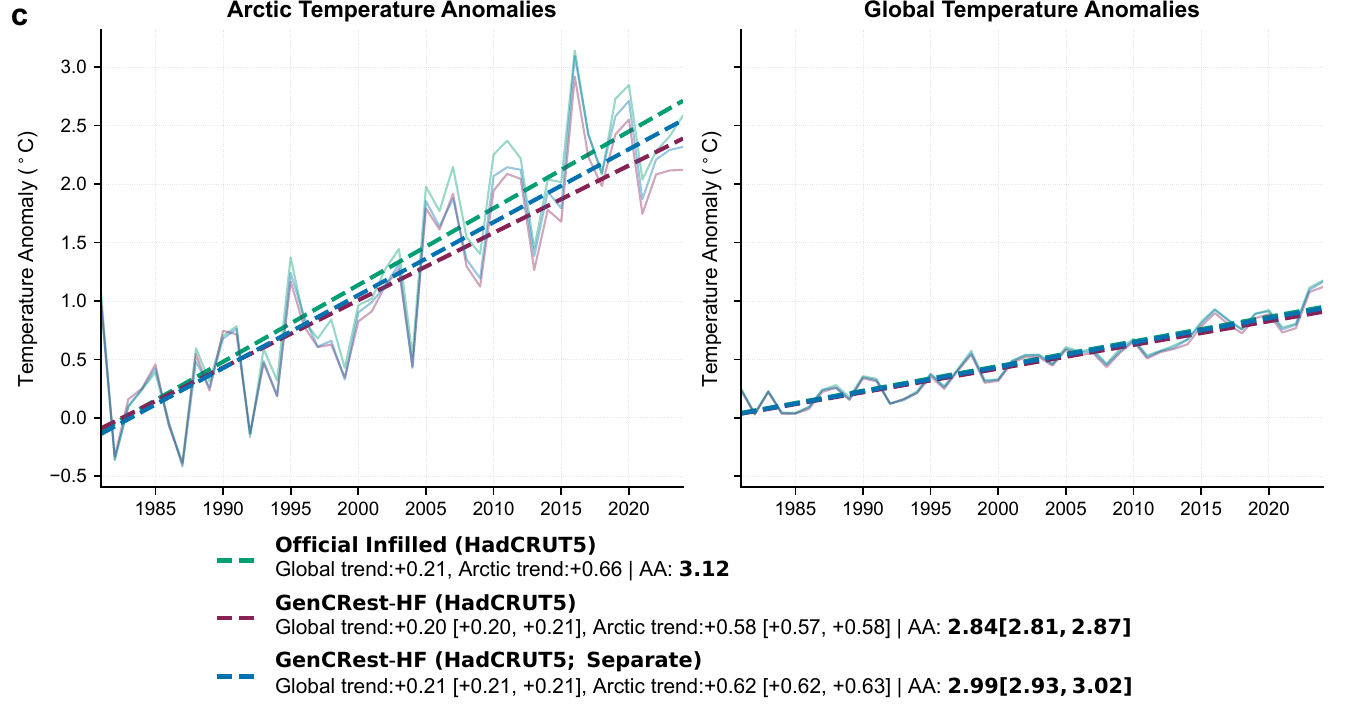}
\caption{
\textbf{Arctic amplification and localized warming in the Barents and Kara Seas.}
Panels compare the HadCRUT5 official infilled product with GenCRest-HF reconstructions using sea-ice-masked inputs or separate air-temperature and sea-surface-temperature models (Supplementary Section~B.6.5).
\textbf{a}, Local amplification during 1981--2024, defined as the grid-cell warming trend divided by the corresponding global-mean trend.
The dashed circle marks $66.5^\circ$\,N; blue and yellow outlines identify the Barents and Kara Seas.
\textbf{b}, Unweighted distributions of grid-cell amplification across the Arctic, Barents Sea, and Kara Sea.
Violins show the spatial distributions; boxes span the interquartile range, with central lines marking the medians.
Whiskers extend to the most extreme values within 1.5 times the interquartile range below or above the box, and points indicate values beyond the whiskers.
\textbf{c}, Area-weighted annual Arctic and global temperature anomalies (solid curves) and fitted trends (dashed lines).
Arctic amplification (AA) is the ratio of Arctic-mean to global-mean warming trends.
Reconstruction maps and time series show means across 10 GenCRest-HF members, with amplification ratios calculated per member before averaging.
Annotations report trends in ${}^\circ\mathrm{C}\,\mathrm{decade}^{-1}$ and AA; bracketed ranges span the minimum and maximum member estimates.
}
\label{fig:temp5_arctic_AA}
\end{figure}

\clearpage
 \begin{figure}[htbp]
\centering
\includegraphics[width=\linewidth]{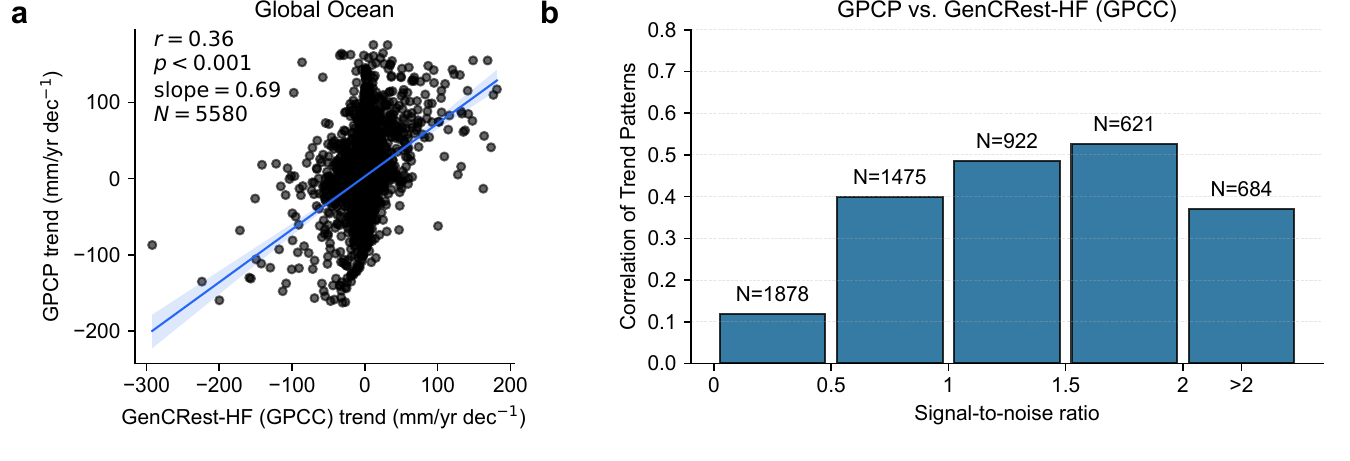}
\includegraphics[width=\linewidth]{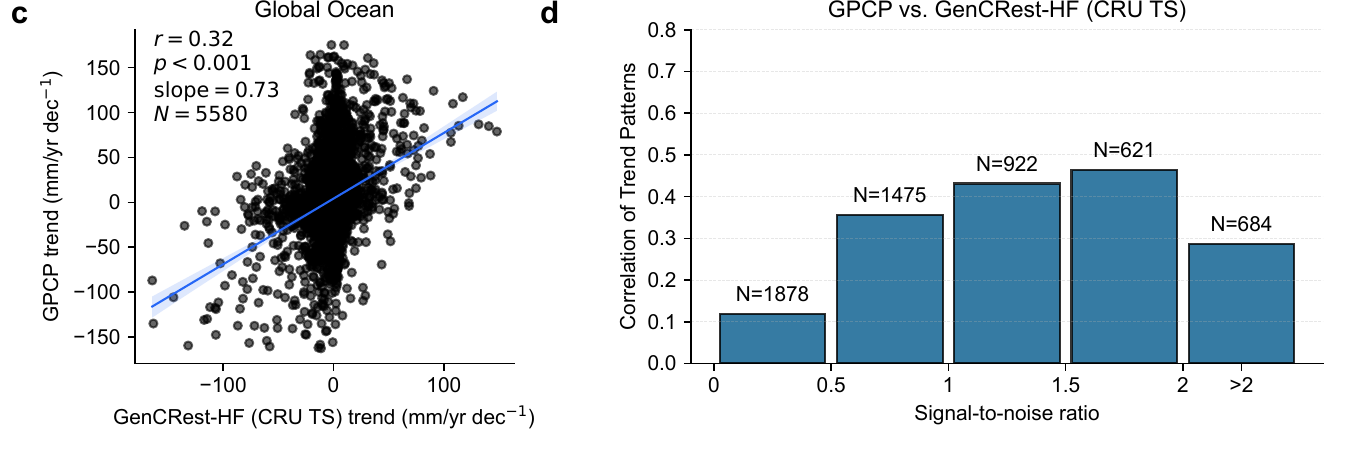}
\includegraphics[width=\linewidth]{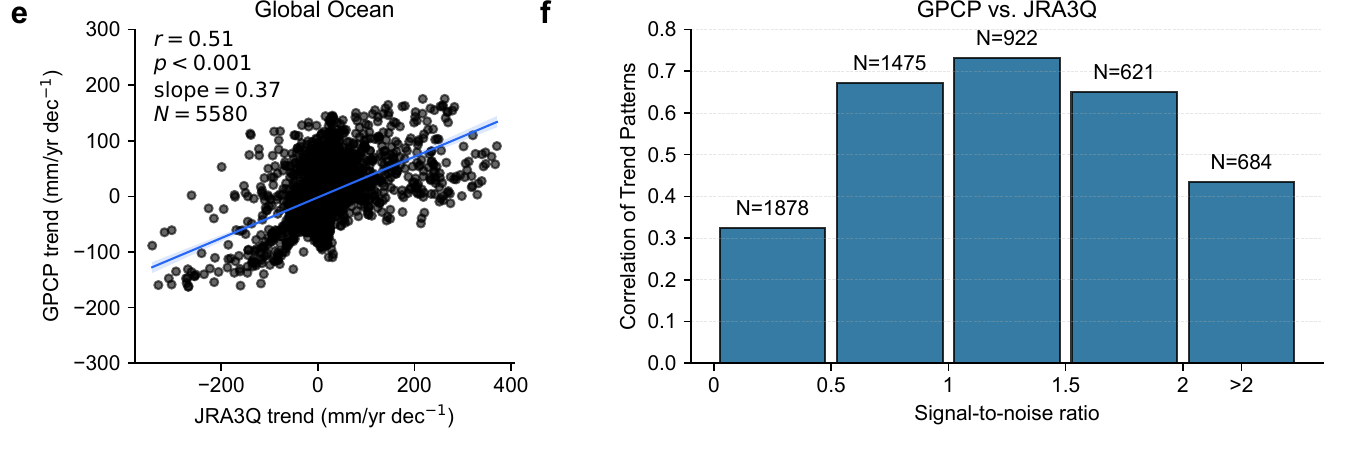}
\caption{
\textbf{Ocean precipitation trends inferred from land observations and their agreement with satellite estimates.}
Linear trends in annual precipitation over 1980--2025 are compared with GPCP v2.3 on a common $2.5^\circ$ grid.
\textbf{a}, \textbf{c}, \textbf{e}, Grid-cell comparisons of GPCP trends with the GPCC v2025 GenCRest-HF reconstruction, the CRU TS 4.10 GenCRest-HF reconstruction, and JRA-3Q, respectively.
Reconstructed trends are calculated from the corresponding 10-member ensemble-mean fields.
Each comparison includes 5,580 ocean grid cells.
Blue lines show ordinary least-squares fits, with shading indicating pointwise 95\% confidence intervals obtained from 1,000 bootstrap resamples.
Annotations report Pearson spatial correlations ($r$), associated $p$ values, regression slopes and sample sizes ($N$).
Trends are expressed in $\mathrm{mm}\,\mathrm{yr}^{-1}\,\mathrm{decade}^{-1}$.
\textbf{b}, \textbf{d}, \textbf{f}, Spatial correlations within bins of the GPCP trend signal-to-noise ratio (SNR) for the corresponding comparisons.
Labels above the bars indicate the number of grid cells in each bin.
All three comparisons yield $p<0.001$.
Correlations are higher outside the lowest-SNR bin, indicating greater agreement where the GPCP trend signal is clearer.
}
\label{fig:ocean_precip_scatter}
\end{figure}

\end{document}